\documentclass[journal]{IEEEtran}

\usepackage{graphicx}
\usepackage{enumerate}
\usepackage{dsfont}
\usepackage[cmex10]{amsmath}
\usepackage{amssymb}
\usepackage{mathrsfs}
\usepackage{algorithm}
\usepackage{algorithmic}
\usepackage{amsmath}
\usepackage{multicol}
\usepackage{stfloats}
\usepackage{array}
\usepackage{cite}
\usepackage{color}
\usepackage{subfigure}
\usepackage{bm}
\usepackage{multirow}

\makeatletter
\newcommand{\Rmnum}[1]{\expandafter\@slowromancap\romannumeral #1@}
\makeatother

\ifCLASSINFOpdf

\else

\fi

\IEEEoverridecommandlockouts

\begin{document}

\title{A Survey of Optimization Approaches for Wireless Physical Layer Security}

\author{Dong Wang, 
        Bo Bai, \IEEEmembership{Senior Member,~IEEE,}
        Wenbo Zhao,
        and Zhu Han, \IEEEmembership{Fellow,~IEEE}


\thanks{The research is partially supported by National Natural Science Foundation of China (NSFC) No. 61871401, Anhui Provincial Natural Science Foundation No. 1708085MF139 and No. 1708085QF158, US AFOSR MURI 18RT0073, NSF CNS-1717454, CNS-1731424, CNS-1702850, CNS-1646607, and ECCS-1547201.}
\thanks{D. Wang is with New Star Research Institute of Applied Technology, Hefei 230031, China (e-mail: eewgdg@yeah.net).} 
\thanks{B. Bai is with Future Network Theory Lab, 2012 Labs, Huawei Technologies Co. Ltd., Hong Kong, China (e-mail: baibo8@huawei.com).}  
\thanks{W. Zhao is with New Star Research Institute of Applied Technology, Hefei 230031, China (e-mail: Zhao-wen-bo@139.com).}
\thanks{Z. Han is with the University of Houston, Houston, TX 77004 USA (e-mail: zhan2@uh.edu $<$mailto: zhan2@uh.edu$>$), and also with the Department of Computer Science and Engineering, Kyung Hee University, Seoul, South Korea.}}

\maketitle

\begin{abstract}

Due to the malicious attacks in wireless networks, physical layer security has attracted increasing concerns from both academia and industry. The research on physical layer security mainly focuses either on the secrecy capacity/achievable secrecy rate/capacity-equivocation region from the perspective of information theory, or on the security designs from the viewpoints of optimization and signal processing. Because of its importance in security designs, the latter research direction is surveyed in a comprehensive way in this paper. The survey begins with typical wiretap channel models to cover common scenarios and systems. The topics on physical-layer security designs are then summarized from resource allocation, beamforming/precoding, and antenna/node selection and cooperation. Based on the aforementioned schemes, the performance metrics and fundamental optimization problems are discussed, which are generally adopted in security designs. Thereafter, the state of the art of optimization approaches on each research topic of physical layer security is reviewed from four categories of optimization problems, such as secrecy rate maximization, secrecy outrage probability minimization, power consumption minimization, and secure energy efficiency maximization. Furthermore, the impacts of channel state information on optimization and design are discussed. Finally, the survey concludes with the observations on potential future directions and open challenges.

\end{abstract}

\begin{IEEEkeywords}

Physical layer security, optimization, resource allocation, beamforming, precoding, cooperative transmission

\end{IEEEkeywords}

%
\IEEEpeerreviewmaketitle

\section{Introduction}

With the rapid evolution of information and communication technologies, mobile Internet and Internet of things (IoT) have become indispensable in daily life. As the foundation of these networks, the cellular network has been designed to support Internet connectivity and full interworking with heterogeneous wireless access networks \cite{CaoMa2014}. This fact, therefore, leads to complicated network architectures, network topologies, access technologies, service requirements, and mobile equipments while bringing serious security issues in wireless information transmission. How to guarantee the security of confidential information has become the precondition to the commercial application of some emerging wireless networks and communication services. Therefore, the theories and technologies of information security have attracted increasing concerns from both academia and industry recently.

\renewcommand\arraystretch{1.5}  
\begin{table*}[t]
\centering
\caption{Comparison between cryptographic encryption and physical layer security}
\label{comparison_tab}
\begin{tabular} { p {4 cm}<{\raggedright}  p{5cm}<{\raggedright}   p{6cm}<{\raggedright}}  
\hline

\hline
& Cryptographic encryption  & Physical layer security  \\
\hline

\hline
  Theoretical basis   & Cryptography   & Information theory  \\
\hline
  Secrecy level     &Can be deciphered by brute-force computing    & Achieving perfect secrecy  \\
\hline
 Computing ability requirements   &Heavily relying on the computing ability       & Being independent of computing ability \\
\hline
 Encryption key management    & Heavy costs resulting from key generation, management, and distribution   &  With no need of any key \\
\hline
 Evaluation criterion   &  Being unable to accurately assess the leakage of confidential information    & Evaluating secrecy precisely by equivocation rate \\
\hline
  Adaptability to channel changes    &  Poor channel adaptability   & Adjusting transmission strategies and parameters to well adapt the channel changes  \\
\hline

\hline
\end{tabular}
\end{table*}

During last few decades, the information security mostly depends on the cryptographic encryption and decryption methods which are deployed at the upper layers of protocol stack. The encryption-based security technologies have been shown to be effective in many cases, but their inherent vulnerabilities are heavy computation and key management costs which may result in high complexity and resource consumption \cite{Stallings2011Security}. As an alternative security technology, the physical layer security, based on the information theory framework, is to utilize the inherent randomness of the physical medium and the difference between the legitimate channels and the wiretap channels to guarantee secure information transmission \cite{Liang2010Information}. Compared with cryptographic approaches, as shown in Table \ref{comparison_tab}, the physical layer security does not rely on the computing capability of the communication equipments, and thus has the advantages of lower complexity and resource savings. It has been shown from the viewpoint of information theory that the physical layer security can achieve perfect secrecy even if the eavesdropper has very strong computing capability. Besides, the physical layer security has a performance metric for secrecy evaluation, i.e., equivocation rate which measures the uncertainty of the confidential message at eavesdroppers. Furthermore, by exploiting the physical layer features, this security technique can flexibly adjust transmission strategies and parameters to accommodate the channel changes. In summary, physical layer security presents distinctive advantages and promising prospects. Therefore, the physical layer security can be used as an effective  supplementary for the cryptographic techniques to further enhance information security.


The concept of secrecy communication was first proposed in the pioneering work of Shannon in 1949 \cite{Shanon1949}, in which secrecy communication was investigated from the viewpoint of information theory. It was proposed therein that the approach termed ``one-time pad" could achieve the perfect secrecy. However, it was very difficult to apply this method in practice due to the intractable difficulties of key generation and management. Being different from the Shannon's model of secrecy communication, Wyner proposed the wiretap channel model in 1975 \cite{Wyner1975wiretap}, in which the perfect secrecy could be achieved at the physical layer by utilizing the difference between the legitimate channel and the illegitimate channel without any key. In Wyner's wiretap channel model, the signal received by the eavesdropper was a degraded version of the signal received by the destination. The characteristic of signal degradation at the eavesdropper made it possible to achieve secrecy at the physical layer. It was also proved by Wyner that the secrecy capacity of a discrete memoryless channel was the maximum value of the difference between the mutual information of the legitimate link and the mutual information of the wiretap link. Thereafter, Csisz$\acute{\text{a}}$r and K$\ddot{\text{o}}$rner generalized the degraded wiretap channel to broadcast channel with confidential messages, and analysed the secrecy capacity of a more general (non-degraded) wiretap channel \cite{Csiszar1978Broadcast}. Following these works, Leung-Yan-Cheong and Hellman investigated the Gaussian wiretap channel and derived the secrecy capacity which is the difference of legitimate channel capacity and wiretap channel capacity \cite{Leung1978Gaussian}. Nevertheless, the early research work cannot be applied directly, since the physical layer security needed suitable secure coding schemes to match the channel states. However, the secure coding technology was less developed in early stage, and the theories and technologies on physical layer security were thus believed to be impractical. Moreover, the fact that the encryption-based security technologies held a dominant position for a long time affected the development of physical layer security. In recent decades, the encryption-based security technologies have exposed some limitations in practical applications. Meanwhile, the coding theories and technologies have got a rapid development, which laid a solid foundation of physical layer security. Accordingly, more and more attentions have been paid on the physical layer security.

The studies on physical layer security can be roughly summarized from two main aspects: 1) the studies related to secrecy rate/capacity from the perspective of information-theoretic security, 2) and the studies related to system designs from the viewpoints of optimization and signal processing \cite{ Raef2013Coop, Hong2013Enhancing, Mukh2014Principles, Yener2015Wireless, Wang2015Enhancing}. The first aspect mainly focuses on the secrecy capacity, achievable secrecy rate, and capacity-equivocation region based on the ideas of information theory. On the other hand, the second aspect mainly focuses on the secure strategy designs based on the techniques of optimization and signal processing. Because of the importance in practical security designs, our objective in this survey is to provide a comprehensive overview on the optimization and design of secure physical layer transmission. The investigations on the topic that we just mentioned are based on the framework of information-theoretic security, since all involved performance metrics, optimization problems, and security solutions in this survey are intertwined with the secrecy rate/capacity which are based on information theory.

\renewcommand\arraystretch{1.5}  
\begin{table*}[t]
\centering
\caption{Brief summaries on existing surveys}
\label{summaries_surveys}
\begin{tabular} { p {1 cm}<{\raggedright}  p{3cm}<{\raggedright}  p{4.3cm}<{\raggedright}   p{7.8cm}<{\raggedright}}  
\hline

\hline
   Surveys   & Publications  &  Focused issues  &   Main contents  \\
\hline

\hline
  \cite{CaoMa2014}    &  IEEE Communications Surveys $\&$ Tutorials    &  Security for long term evolution (LTE) and LTE-advanced Networks.  &  Security functionalities, security vulnerabilities, and existing security solutions.  \\ 
\hline
  \cite{Mukh2014Principles} & IEEE Communications Surveys $\&$ Tutorials   & Physical layer security in multiuser wireless networks. &  Security improvements in multi-antenna, broadcast, multiple-access, interference, and relay channels, as well as physical-layer key generation and secure coding.   \\ 
\hline
  \cite{Liu2016Phynext}  &  IEEE Communications Surveys $\&$ Tutorials   & Comprehensive overview on the fundamentals and technologies of physical layer security.   &    Technologies, challenges, and solutions in physical layer security are studied from the aspects of wiretap coding, multi-antenna and relay cooperation, physical-layer key generation, and physical-layer authentication.     \\ 
\hline
  \cite{Yener2015Wireless}  & Proceedings of the IEEE  &   Lessons learned from information-theoretic security with multiple
wireless transmitters.  & Designing secure wireless systems with unauthenticated entities by cooperative jamming/relaying and interference alignment. \\ 
\hline
  \cite{Zou2016security}   & Proceedings of the IEEE  &  Security vulnerabilities, security threats, and efficient defense mechanisms.   &  Discussing the security requirements and attacks at each protocol layer, investigating the existing security protocols and algorithms, while exploring the state of the art in physical layer security. \\   
\hline
  \cite{Mukherjee2015PhyIoT}   & Proceedings of the IEEE  & Physical layer security in the Internet of Things.  & Surveying the advances and challenges in resource constrained secrecy coding and secret-key generation in the Internet of Things.  \\  
\hline
  \cite{Raef2013Coop}  & IEEE Signal Processing Magazine   & Cooperative security at physical layer.  & Guaranteeing information security by using cooperative techniques which consist of carefully designed coding and signaling schemes. \\  
\hline
  \cite{Hong2013Enhancing}  &  IEEE Signal Processing Magazine  & Signal processing techniques for secrecy in multi-antenna wireless systems.   & Enhancing physical layer security in multi-antenna systems  by  beamforming/precoding with or without artificial noise. \\  
\hline
  \cite{Wang2015Enhancing}  &  IEEE Communications Magazine    &  Recent research on enhancing secrecy via cooperation.    &  Signal design and optimization to increase secrecy based on cooperative relaying and jamming.     \\ 
\hline
  \cite{Zhou2014Jointmult}  &  IEEE Communications Magazine    & A joint framework involving both the physical layer and application layer security technologies.   & Proposing a joint security scheme  by exploiting the security capacity and signal processing technologies at the physical layer and the authentication and watermarking strategies at the application layer.    \\   
\hline
  \cite{Kapetanovic2015Physical} & IEEE Communications Magazine    &   Physical layer security for massive MIMO.  & Discussing the passive eavesdropping and active attacks in massive MIMO systems while proposing three detection schemes to identify the active attacks.     \\ 
\hline
  \cite{Rod2015Physical}   & IEEE Communications Magazine    &  Physical layer security in cooperative relay networks.   &   Pure or hybrid relaying/jamming combinations for secrecy improvements with trusted/untrusted relays.   \\  
\hline
  \cite{Yang2015Safeguarding} &  IEEE Communications Magazine &  Physical layer security in the 5G network. &  The opportunities and challenges offered by the disruptive technologies enabling 5G for achieving high physical layer security.  \\   
\hline
  \cite{WadeTrappe2015Cha}   &  IEEE Communications Magazine &  Challenges of physical layer security in practical applications. &  Identifying the important issues to apply physical layer security into practice.  \\  
\hline
  \cite{Shiu2011Physicaltut}  & IEEE Wireless Communications & Several prevalent methods to enhance physical layer security.   &  Classifying the methods of physical layer security into five major categories while comparing their reliability, computational complexity, and secrecy capacity.     \\   
\hline
  \cite{Zou2014Improving}  & IEEE Network  & Diversity techniques to improve physical layer security.    & Exploiting MIMO diversity, multiuser diversity, and
cooperative diversity to secure wireless communications.  \\    
\hline

\hline
\end{tabular}
\end{table*}

Many excellent surveys have been published in physical layer security, which provide comprehensive overviews and insightful comments to understand the fundamental principles, technology status, and future trends in this field. In \cite{Liu2016Phynext}, the fundamentals and technologies of physical layer security are reviewed comprehensively. Specifically, in \cite{Liu2016Phynext}, the technologies, challenges, and solutions are summarized from more methodological viewpoints involving wiretap coding, multi-antenna and relay cooperation, physical-layer key generation, and physical-layer authentication. Moreover, we highlight the focused issues and the main contents of some published surveys in Table \ref{summaries_surveys}. In contrast to existing surveys, our work tries to review the recent advances in physical layer security from the perspective of system optimization and design. First, we summarize the research topics and the secure strategies that cover extensive problems in system optimization and design, such as secure resource allocation, signal processing and cooperative diversity. Second, the performance metrics and the related optimization problem formulations are investigated to provide deep insights into secure transmission designs. Finally, we survey the state of the art of optimization and design on each research topic of physical layer security from four categories of basic optimization problems, i.e., maximization of achievable secrecy rate, minimization of secrecy outrage probability, minimization of power consumption, and maximization of secure energy efficiency (EE). In particular, some optimization approaches and secure strategies which are usually appeared in physical-layer transmission designs are summarized with detailed procedures.

In summary, this survey provides a well-rounded overview for newcomers to understand the optimization and design in physical layer security. The contributions of this survey is based on the following work: 1) Summarizing general wiretap channel models to cover the basic scenarios in this field, followed with usually appeared optimization approaches. 2) Investigating hot topics in physical layer security from the perspective of system optimization and design. 3) Seeking deep insights into performance metrics to achieve different requirements in system designs. 4) Reviewing the state of the art of optimization and design in this field and the harmful impacts of channel state information (CSI) on designing security solutions. 5) Discussing future possible directions and open challenges.

\renewcommand\arraystretch{1.5}  
\begin{table}[t]
\centering
\caption{Abbreviations and their definitions}
\label{Table_Abbr}
\begin{tabular} { p {1.5 cm}<{\raggedright}  p{5.5cm}<{\raggedright}   }  
\hline

\hline
Abbreviation   &   Definition    \\
\hline

\hline
  5G      &  The fifth generation \\
\hline
  AF      &  Amplify-and-forward   \\
\hline
  AN      &  Artificial noise     \\
\hline
  CSI     &  Channel state information \\
\hline
 CI       &  Channel inversion\\
\hline
  DC      &  Difference of convex functions  \\
\hline
  DF      &  Decode-and-forward    \\
\hline
  EE      &  Energy efficiency     \\
\hline
  GSVD    &  Generalized singular value decomposition \\
\hline
  I/Q     & In-phase and quadrature    \\
\hline
  IoT     &  Internet of things  \\
\hline
  LTE     &  Long term evolution  \\
\hline
  MIMO    &  Multiple-input multiple-output \\
\hline
  MIMOME  &  Multi-input multi-output multi-eavesdropper \\
\hline
  MISO    &  Multiple-input single-output   \\
\hline
  MISOME  &  Multi-input single-output multi-eavesdropper   \\
\hline
  MRC     &  Maximal ratio combining   \\
\hline
  MRT     &  Maximum ratio transmission      \\
\hline
  mm-Wave &  Millimeter-wave    \\
\hline
  OFDMA   &  Orthogonal frequency division multiple access \\
\hline
  QoS     &  Quality of service      \\
\hline
  RCI     &  Regularized channel inversion   \\
\hline
  SISO    &  Single-input single-output     \\
\hline
  SIMO    &  Single-input multiple-output   \\
\hline
  SDP     &  Semidefinite programming  \\
\hline
  SDR     &  Semidefinite relaxation\\
\hline
  SE      &  Spectrum efficiency     \\
\hline
  SPCA    &  Sequential parametric convex approximation   \\
\hline
  SNR     &  Signal-to-noise ratio   \\
\hline
  SINR    &  Signal-to-interference-plus-noise ratio   \\
\hline
  S-DPC   &  Secret dirty-paper coding   \\
\hline
  ZF      &  Zero-forcing  \\
\hline

\hline
\end{tabular}
\end{table}

\begin{figure}[ht]
\centering
\includegraphics[width=3.5in]{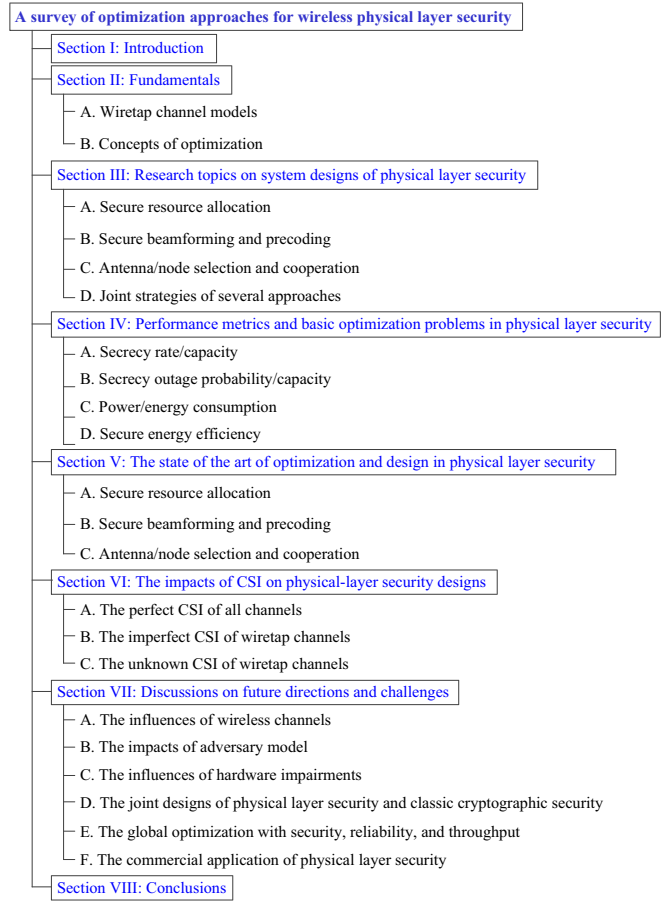}  
\caption{The structural diagram of this survey.}
\label{Structural_diagram}
\end{figure}

The remainder of this paper is organized as follows. In Section \ref{section_sys_model}, typical wiretap channel models and optimization concepts are introduced  to cover common communication scenarios and optimization approaches. In Section \ref{section_secure_topics}, the research topics in physical layer security are investigated from the perspective of secure resource allocation, beamforming/procoding, and antenna/node selection and cooperation. In Section \ref{section_metrics}, we seek deep insights into performance metrics which can be adopted in all research topics to evaluate the proposed secure transmission strategies. The state of the art of optimization and design in physical layer security is reviewed in Section \ref{section_state_art}, followed with usually appeared optimization approaches and security strategies. Section \ref{section_CSI} investigates the common assumptions of CSI and their negative impacts on secure transmission designs. Future possible directions and open challenges are discussed In Section \ref{section_Direction} to provide some lessons for newcomers. Finally, the survey is concluded in Section \ref{section_conclusion}. A diagram is illustrated in Fig. \ref{Structural_diagram} to show the outline and structure of this paper. In addition, abbreviations used in this paper are defined in Table \ref{Table_Abbr}.

\emph{Notations}: Throughout this paper, matrices and vectors are denoted by bold uppercase letters and bold lowercase letters, respectively. $\bm{x}$ denote the set of optimization variable without physical meaning. Mutual information, conjugate transpose, and Euclidean norm are represented by $I(\cdot ;  \cdot)$, $(\cdot)^H$, and $\| \cdot \|$, respectively. The trace of a matrix is denoted by $\text{Tr}()$. $\textbf{W} \succeq \textbf{0}$  means that $\textbf{W}$ is a positive semidefinite matrix.

\section{Fundamentals} \label{section_sys_model}

In this section, we give several typical wiretap channel models to cover the common scenarios and systems considered in the survey, and introduce general concepts of optimization and optimization problems to clarify the variables and parameters in security designs.

\subsection{Wiretap Channel Models}

The typical wiretap channel models usually include multiple-input multiple-output (MIMO) wiretap channels, broadcast wiretap channels, multiple-access wiretap channels, interference wiretap channels, and relay wiretap channels \cite{Mukh2014Principles}, etc.

\subsubsection{MIMO wiretap channels}

The simplest network in physical layer security is composed of a transmitter, a legitimate receiver, and an unauthorized receiver (eavesdropper), in which confidential messages are exchanged between the transmitter and the legitimate receiver while protecting from the unauthorized receiver. In such a scenario, the terminals may be equipped with multiple antennas. The typical channel model for multi-antenna scenarios is the MIMO wiretap channel which can cover the special models of single-input single-output (SISO), single-input multiple-output (SIMO), and multiple-input single-output (MISO) channels. In the MIMO channel in which the transmitter, receiver, and eavesdropper are deployed with $n_t$, $n_d$, and $n_e$ antennas, respectively, the general expressions for the received signals at the legitimate receiver and eavesdropper are, respectively, given by \cite{Mukh2014Principles}
    \begin{equation}
    \label{MIMO_rec_signal}
        \textbf{y}_d  =  \textbf{H}_d \textbf{x}_{s} + \textbf{z}_d,
    \end{equation}
    \begin{equation}
    \label{MIMO_rec_eve}
        \textbf{y}_e  =  \textbf{H}_e \textbf{x}_s + \textbf{z}_e,
    \end{equation}
where $\textbf{x}_s$ is the ${n_t \times 1}$ encoded signal with a covariance matrix constraint $\mathds{E}\{\textbf{x}_s\textbf{x}_s^H\}=\textbf{Q}_x $ for $\textbf{Q}_x\succeq \textbf{0}$ or an average power constraint $\text{Tr}\{\textbf{Q}_x\} \le P_{max}$ for a peak power $P_{max}$. The ${n_d \times n_t}$ matrix $\textbf{H}_d$ and the $n_e \times n_t$ matrix $\textbf{H}_e$ are the channel gain matrices to the legitimate receiver and the eavesdropper, respectively. $\textbf{z}_d$ and $\textbf{z}_e $ are white Gaussian noise vectors at the legitimate receiver and eavesdropper, respectively. This wiretap channel model is typical, and to be widely investigated in physical layer security.

\subsubsection{Broadcast wiretap channels}


The broadcast wiretap channels are raised in multi-user networks with more than two receivers where one transmitter delivers confidential information to multiple users with the presence of multiple eavesdroppers. We assume that there are one transmitter equipped with $n_t$ antennas, $I$ users each with $n_{d_i}$ antennas, and $J$ eavesdroppers each with $n_{e_j }$ antennas. In the downlink, the transmitter transmits confidential messages to the legitimate users while preventing from overhearing of the eavesdroppers. This broadcast channel can be equivalent to a compound wiretap channel which is defined as \cite{Liang2010Information, Ekrem2012Secure}
    \begin{equation}
    \label{rec_BC}
        \textbf{y}_{d_i}  =  \textbf{H}_{d_i}  \textbf{x}_s + \textbf{z}_{d_i}, i = 1,2,\cdots, I,
    \end{equation}
    \begin{equation}
    \label{rec_BC}
        \textbf{y}_{e_j}  =  \textbf{H}_{e_j}  \textbf{x}_s + \textbf{z}_{e_j}, j = 1,2,\cdots, J,
    \end{equation}
where $\textbf{x}_s$ denotes the $n_t \times 1$ encoded signal for the confidential messages which is subject to a covariance matrix constraint $\mathds{E}\{\textbf{x}_s\textbf{x}_s^H\}=\textbf{Q}_x $ for $\textbf{Q}_x\succeq \textbf{0}$ or an average power constraint $\text{Tr}\{\textbf{Q}_x\} \le P_{max}$ for a peak power $P_{max}$. $\textbf{y}_{d_i}$ and $\textbf{y}_{e_j}$ are the received signals at user $i$ and eavesdropper $j$, respectively. $\textbf{H}_{d_i}$ is $n_{d_i} \times n_t$ channel matrix to user $i$ and $ \textbf{H}_{e_j} $ is $n_{e_j} \times n_t$ channel matrix to eavesdropper $j$. $\textbf{z}_{d_i}$ and $\textbf{z}_{e_j}$ are white Gaussian noise vectors at user $i$ and eavesdropper $j$, respectively. The compound wiretap channel has several special cases including the parallel wiretap channel with two eavesdroppers, the fading wiretap channel with multiple eavesdroppers, and the wiretap channel with multiple receivers \cite{Liang2010Information}, etc. In addition, another specific broadcast channel is the broadcast channel with separate confidential messages of each user in which each downlink message must be kept secret from all other unintended users (each user is seen as an eavesdropper for messages not intended to it) \cite{Mukh2014Principles}.


\subsubsection{Multiple-access wiretap channels}


In the multiple-access wiretap channel, multiple transmitters transmit messages to a legitimate receiver with the existence of an eavesdropper. There are $K$ transmitters each with $n_{t_k}$ antennas, one legitimate receiver with $n_{d}$ antennas, and one eavesdropper with $n_e$ antennas. Let us define $n_{d} \times n_{t_k}$ matrix $\textbf{H}_{d_k}$ and $n_{e} \times n_{t_k}$ matrix $\textbf{H}_{e_k}$ as the channel matrices from transmitter $k$ to the receiver and the eavesdropper, respectively. Then, the received signals at the receiver and the eavesdropper are, respectively, expressed as \cite{Lee2017PrecoderMAC}
    \begin{equation}
    \label{rec_MAC}
        \textbf{y}_d  = \sum\limits_{k=1}^K \textbf{H}_{d_k} \textbf{x}_{s_k} + \textbf{z}_d,
    \end{equation}
    \begin{equation}
    \label{rec_MAC}
        \textbf{y}_e  = \sum\limits_{k=1}^K \textbf{H}_{e_k} \textbf{x}_{s_k} + \textbf{z}_e,
    \end{equation}
where $\textbf{x}_{s_k}$ denotes the $n_{t_k} \times 1$ encoded signal at transmitter $k$ with a covariance matrix constraint or an average power constraint. $\textbf{z}_{d}$ and $\textbf{z}_{e}$ are the white Gaussian noise vectors at the receiver and the eavesdropper, respectively. Some special cases of multiple-access channel in physical layer security are also investigated, such as SISO multiple-access channel with an eavesdropper \cite{Tekin2008TheMAC} and multiple-access channel with common and confidential messages \cite{Liang2010Information}.


\subsubsection{Interference wiretap channels}


The interference wiretap channel refers to the scenario where multiple links are simultaneously active in the same time and frequency slot, and hence potentially interfere with each other \cite{Fakoorian2011MIMO}. At the same time, the communications over the multiple links are overheard by an eavesdropper. We consider the interference wiretap channel with $K$ user pairs and an eavesdropper, where the source user $k$, the destination user $k$, and the eavesdropper are deployed with $n_{t_k}$, $n_{d_k}$, and $n_e$ antennas, respectively, $k=1, \cdots, K$. The received signals of destination user $k$ and the eavesdropper are, respectively, written as \cite{Kong2016Iterative}
    \begin{equation}
    \label{rec_MAC}
        \textbf{y}_{d_k}  = \textbf{H}_{d_{kk}} \textbf{x}_{s_k} +  \sum\limits_{l \neq k}^K \textbf{H}_{d_{kl}} \textbf{x}_{s_l} + \textbf{z}_{d_k},
    \end{equation}
    \begin{equation}
    \label{rec_MAC}
        \textbf{y}_e  = \sum\limits_{l = 1}^K \textbf{H}_{e_l} \textbf{x}_{s_l} + \textbf{z}_{e},
    \end{equation}
where $\textbf{x}_{s_l}$ is the $n_{t_l} \times 1$ transmitted signal of source user $l$ with a covariance matrix constraint or an average power constraint. The $n_{d_k} \times n_{t_l} $ matrix $\textbf{H}_{d_{kl}}$ denotes the channel matrix from source user $l$ to destination user $k$. The $n_e \times n_{t_l} $ matrix $\textbf{H}_{e_{l}}$ denotes the channel matrix from source user $l$ to the eavesdropper.  $\textbf{z}_{d_k}$ and $\textbf{z}_{e}$ are the white Gaussian noise vectors at destination user $k$ and the eavesdropper, respectively. A further model of interest is the interference channel with separate confidential messages, in which each source message must be kept confidential from all other unintended users. A specific case of this channel model is studied in \cite{Liang2010Information} where SISO interference channel is used to deliver two confidential messages.

\subsubsection{Relay wiretap channels}

A typical cooperative wireless network considering physical layer security is consist of a source, a destination, a relay, and an eavesdropper, each with $n_t$, $n_d$, $n_r$, and $n_e$ antennas, respectively. The relay is operated in a decode-and-forward (DF) mode. In the first phase, the source transmits the $n_t \times 1$ signal vector $ \textbf{x}_s$ to the relay. The relay, the destination, and the eavesdropper receive the signal as \cite{Huang2011Cooperative}
    \begin{equation}
    \label{rec_MAC}
        \textbf{y}_{r}  = \textbf{H}_{sr} \textbf{x}_s + \textbf{z}_{r},
    \end{equation}
    \begin{equation}
    \label{rec_MAC}
        \textbf{y}_d^{(1)}  = \textbf{H}_{sd} \textbf{x}_s + \textbf{z}_d,
    \end{equation}
    \begin{equation}
    \label{rec_MAC}
        \textbf{y}_e^{(1)}  = \textbf{H}_{se} \textbf{x}_s + \textbf{z}_e,
    \end{equation}
where the $n_r \times n_t$ matrix $\textbf{H}_{sr}$, the $n_d \times n_t$ matrix $\textbf{H}_{sd}$, and the $n_e \times n_t$ matrix  $\textbf{H}_{se}$ are the channel matrices from the source to the relay, the destination, and the eavesdropper, respectively. $\textbf{z}_{r}$, $\textbf{z}_d$, and $\textbf{z}_e$ are the white Gaussian noise vectors at the relay, the destination, and the eavesdropper, respectively. The relay decodes the received signal and forwards it to the destination. Let the $n_d \times n_{r}$ matrix  $\textbf{H}_{rd}$ and the $n_e \times n_{r}$ matrix $\textbf{H}_{re}$ denote the channel matrices from the relay to the destination and the eavesdropper, respectively. In the second phase, the $n_{r}\times 1$ transmitted signal vector $\textbf{x}_{r}$ of the relay is a new version of $\textbf{x}_s$ by using an encoding scheme. Then, the received signals at the destination and the eavesdropper are, respectively, obtained as
    \begin{equation}
    \label{rec_MAC}
        \textbf{y}_d^{(2)}  =  \textbf{H}_{rd} \textbf{x}_r + \textbf{z}_d,     
    \end{equation}
    \begin{equation}
    \label{rec_MAC}
        \textbf{y}_e^{(2)}  =  \textbf{H}_{re} \textbf{x}_r + \textbf{z}_e.     
    \end{equation}
The other typical cooperative channel model is the amplify-and-forward (AF) relay channel which is also investigated extensively in physical layer security, such as in \cite{Mo2014Secure, Li2015RobustCoop}.

\subsection{Concepts of Optimization}

In this subsection, the concepts of optimization and optimization problems are introduced for understanding the survey easily.

\subsubsection{General optimization problem}

 A general mathematical optimization problem can be formulated as \cite{boyd2004Convex}
    \begin{equation}
    \label{opt_problem}
    \begin{array}{l}
        \mathop{\min \limits_{ \bm{x} } ~ f(\bm{x}) } \\
        \text{s.t.}~~
        \left\{\begin{array}{l}
              h_i (\bm{x}) \le b_i, i = 1,2,\cdots, m,   \\
              g_j (\bm{x}) = c_j, i = 1,2,\cdots, n,
        \end{array}\right.
    \end{array}
    \end{equation}
where $\bm{x}$ is the set of optimization variable. The function $f(\bm{x})$ is the objective function. The constraint conditions $h_i (\bm{x}) \le b_i$ and $ g_j (\bm{x}) = c_j$ are the inequality and equality constraints, respectively. If there is no constraint, we say the problem is unconstrained. The optimization problem formulated in \eqref{opt_problem} describes the problem of finding an optimal $\bm{x}^*$ that minimizes $f(\bm{x})$ among all $\bm{x}$ satisfying the constraints $h_i (\bm{x}) \le b_i$ and $ g_j (\bm{x}) = c_j$. Therefore, $\bm{x}^*$ is called the optimal solution of the problem \eqref{opt_problem}.

Convex optimization is an important class of optimization problem. The standard convex optimization is defined as \cite{boyd2004Convex}
    \begin{equation}
    \label{Convex_problem}
    \begin{array}{l}
        \mathop{\min \limits_{ \bm{x} } ~ f(\bm{x}) } \\
        \text{s.t.}~~
        \left\{\begin{array}{l}
              h_i (\bm{x}) \le b_i, i = 1,2,\cdots, m,   \\
              d_j^T \bm{x} =  c_j, j = 1,2,\cdots, n,
        \end{array}\right.
    \end{array}
    \end{equation}
where $f(\bm{x})$ and $h_i (\bm{x})$ are convex functions. Comparing to problem \eqref{opt_problem}, the convex problem has the characteristics that the objective function and inequality constraint functions must be convex while the equality constraint functions $g_j (\bm{x}) = d_j^T \bm{x} - c_j$ must be affine \cite{boyd2004Convex}. Convex optimization problems can be solved optimally by many efficient algorithms, such as interior-point methods. If a practical problem can be formulated as a convex optimization problem, the original problem can then be solved. Therefore, many problems can be solved via convex optimization by transforming the original problem into a convex optimization problem.

Another class of optimization problem is nonconvex optimization which covers the problems with nonconvex objective function or/and nonconvex constraint functions. The nonconvex optimization problems are usually intractable. The complexity of global optimization methods for nonconvex problems may grow exponentially with the problem sizes. However, some nonconvex problems can be transformed into or approximated by convex problems. By solving the resulting convex problems, we can get the optimal solution of the original nonconvex problems. Moreover, to overcome the difficulties of solving nonconvex problems, some heuristic algorithms can be designed based on convex optimization, such as randomized algorithms in which an approximate solution to a nonconvex problem is found by drawing some number of candidates from a probability distribution, and taking the best one found as the approximate solution \cite{boyd2004Convex}. In addition, for nonconvex problems, the compromise is to give up seeking the optimal solution. Instead, we seek a locally optimal solution by combining convex optimization with a local optimization method, where convex optimization can be used for initialization of local optimization.

\subsubsection{Optimization in physical layer security}

Following the great progress in theories and algorithms of optimization, the system designs in physical layer security has greatly benefited from recent advances to the point where optimization has now emerged as a major signal processing technique.

Towards general optimization problem \eqref{opt_problem} in physical layer security, the objective function $f(\bm{x})$ may be the considered performance metrics, such as secrecy rate/capacity, secrecy outage probability/capacity, power consumption, and secure EE which will be elaborated in Section \ref{section_metrics}. The optimization variable $\bm{x}$ may be the resources in the designs of secure resource allocation, beamformer/precoder in the designs of secure beamforming/precoding, or candidates of antennas/cooperative nodes in the designs of antenna/node selection and cooperation. The secure resource allocation, beamforming/precoding, and antenna/node selection and cooperation mentioned here are the research topics in physical layer security, which will be discussed in detail in Section \ref{section_secure_topics}.

\begin{figure*}[ht]
\centering
\includegraphics[width=5.5in]{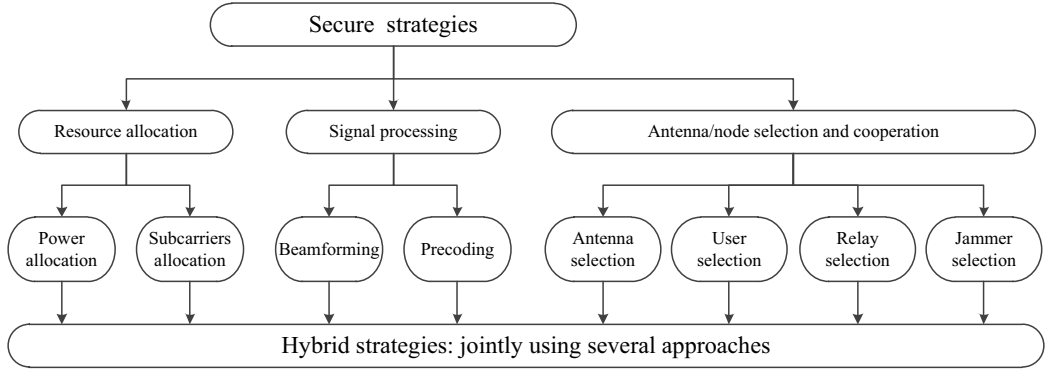}
\caption{Secure strategies for improving physical layer security. }
\label{Summary_secure_strategies}
\end{figure*}

In physical layer security, the majority of optimization problems are nonconvex due to the property of logarithmic subtraction in secrecy rate/capacity. We can roughly list several optimization problems usually appeared in this field as follows.

\begin{itemize}
  \item Integer programming in which some or all optimization variables are constrained to be integer values. This kind of problems is usually raised in the designs of secure subcarrier allocation and antenna/node selection.
  \item Mixed integer programming that concerns the problems having discrete and continuous variables. In joint subcarrier and power allocation, or joint antenna/node selection and beamforming, such problems are dealt with usually.
  \item Difference of convex functions (DC) programming where the objective function is a subtraction of two convex functions. This feature fits with the definition of secrecy rate/capacity. Therefore, DC programming is widely used for solving the problems of secrecy rate maximization.
  \item Quadratic programming where the objective function has quadratic terms. This problem appears in the designs of secure power allocation and beamforming, such as the typical optimization problem of power minimization.
  \item Semidefinite programming (SDP) which optimizes a linear function of the variables subject to linear equality constraints and a nonnegativity constraint on the variables. In physical layer security, some nonconvex problems are usually transformed into SDP to get an efficient algorithm that is easy to implement.
  \item Fractional programming which focuses on optimizing a ratio of two nonlinear functions. The typical example is EE maximization with the considerations of physical layer security.
  \end{itemize}

To cope with the nonconvexity of the optimization problems in physical layer security designs, many optimization techniques have been proposed, such as dual decomposition, alternating search, penalty function method, sequential parametric convex approximation (SPCA), semidefinite relaxation (SDR), and so on, which will be discussed in Section \ref{section_state_art}.

\section{Research Topics on System Designs of Physical Layer Security}\label{section_secure_topics}

Many conventional physical layer technologies of wireless communications without secrecy consideration can be redesigned for confidential information transmission under the framework of physical layer security. From the perspective of system designs, the research topics on physical layer security mainly focus on secure resource allocation, secure beamforming/procoding, secure antenna/node selection\footnote{Node selection usually adopted in multi-node scenarios includes user selection, relay selection, and jammer selection.} and cooperation, and the joint considerations based on the aforementioned strategies, as shown in Fig. \ref{Summary_secure_strategies}.

\subsection{Secure Resource Allocation}

\begin{figure}[t]
\centering
\includegraphics[width=\linewidth]{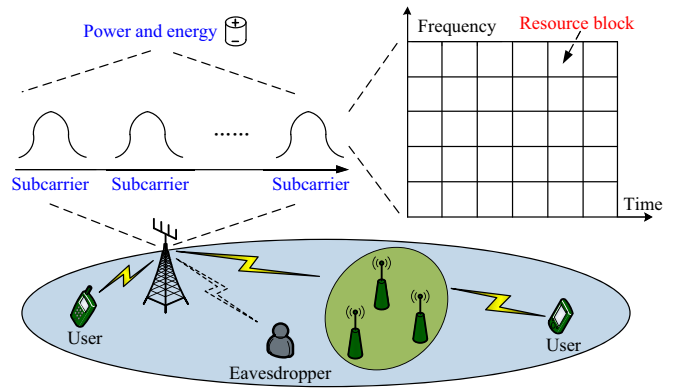} 
\caption{An illustration of the multidimensional wireless resources in a multi-antenna multi-node OFDMA-based wireless network.}
\label{Fig_resource_allocation}
\end{figure}

Resource allocation which has been widely used in the conventional communications without the consideration of secrecy \cite{Guocong2005Utility}, is also an effective way for enhancing physical layer security. The multidimensional wireless resources make it possible to intentionally extend the difference between the legitimate channel and the wiretap channel by secure resource allocation. The multidimensional wireless resources typically contain the frequency, timeslot, and power in orthogonal frequency division multiple access (OFDMA) networks. In multi-antenna and multi-node wireless networks, the wireless resources generally refer to the spatial degrees of freedom provided by multiple antennas and nodes, as shown in Fig. \ref{Fig_resource_allocation}.

Given the limited network resources such as bandwidth and energy, the main challenge of secure resource allocation is to utilize the limited resources as efficient as possible to achieve the requirements of some performance metrics, such as secrecy rate, secrecy outage probability, power consumption, and secure EE. Hence, many works have focused on the two basic problems of secure resource allocation that are the subcarrier allocation and the power allocation in multicarrier networks.

The subcarrier allocation aims at finding the optimal subcarrier usage policy that is able to effectively improve spectral efficiency and information security. Without loss of generality, the secure subcarrier allocation is usually formulated as a binary integer programming \cite{Bo2014Outage, Jindal2015Resource, Tao2011Subcarrier, Ng2011secure}. More specifically, whether or not a subcarrier is used for communication is specified by a decision variable $\alpha \in \{0, 1\}$, with $\alpha =1$ meaning that the subcarrier is used for transmitting and $\alpha = 0$ otherwise.

Adaptive power allocation among multiple carriers and nodes is another important method, which can be applied for a further performance improvement \cite{Krikidis2010Power, Jeong2011Optimal, Huang2012Robust, Park2013On, Wang2014Secure, Poor2014Power, Lee2015Full, Lee2016Optimal, Zheng2015Optimal, Ahmed2016Power, Zhang2016Secrecy, Chen2016Resource}. Accordingly, different strategies based on joint subcarrier and power allocation have been proposed to achieve different design requirements in physical layer security \cite{Ng2011secure, Jindal2015Resource, Tao2011Subcarrier, zhang2015commun, Karach2015Security, Ng2012Energy}. The joint subcarrier and power allocation are generally modeled as mixed integer nonlinear optimization  which is an NP-hard problem in most situations. In practice, a number of optimization techniques have been proposed to provide simple and suboptimal solutions for such combinatorial optimization problems \cite{Tao2011Subcarrier, Jindal2015Resource, Jeong2011Optimal, Karach2015Security, zhang2015commun}.

\subsection{Secure Beamforming and Precoding}

Signal processing techniques, such as beamforming and precoding which are popular in  multi-antenna and multi-node cooperative networks, have been demonstrated as promising ways to achieve the physical layer security \cite{Hong2013Enhancing}. The deployment of multi-antenna and multi-node cooperative networks is thought to have great potential to enhance not only transmission effectiveness and reliability but also wireless security. It has been verified that collaborative beamforming and precoding in multi-antenna and multi-node cooperative networks can bring some benefits in terms of the secrecy rate, secrecy outage probability, power/energy consumption, and secure EE.

\begin{figure}[t]
\centering
\includegraphics[width=0.7\linewidth]{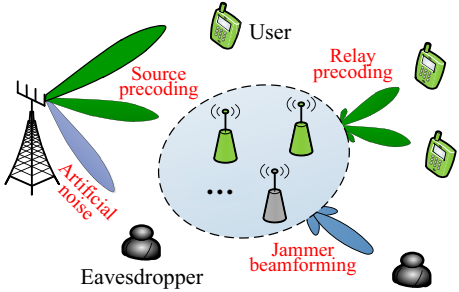}
\caption{An illustration of secure beamforming and precoding in a multi-antenna and multi-node cooperative network.}
\label{Fig_beamforming_precoding}
\end{figure}

Beamforming and precoding technologies have been exploited to achieve different performance requirements in secure transmission. Secure beamforming typically refers to one-rank transmission by which only single data stream is transmitted over multiple antennas or nodes, whereas secure precoding refers to multi-rank transmission by which more than one data streams can be transmitted at the same time \cite{Hong2013Enhancing}. Generally speaking, beamforming serves as a special case of precoding. An illustration of secure beamforming and precoding in multi-antenna and multi-node cooperative networks is shown in Fig. \ref{Fig_beamforming_precoding}, where the source precoding is assisted by artificial noise (AN) and the intermediate nodes are used for relay precoding and jammer beamforming.

The main idea of secure beaforming is to compute the optimal beamforming vector for achieving some performance metrics of physical layer security by enhancing the signal quality at the destination node and decreasing the signal quality at the eavesdropper. Most of the secure beamforming involves solving optimization problems. Due to the special form of logarithmic subtraction in the secrecy rate, the optimization problems of secure beamforming are usually neither convex nor concave in many situations. Therefore, they can only be solved by numerical methods with high complexity, such as in \cite{Jeong2012Joint, Mo2014Secure, Liu2014Transmit, Shi2014Secure, Lv2015Secrecy, Zhao2015Robust}. To mitigate the computational cost, some low-complexity suboptimal algorithms have been proposed to simplify the beamforming designs \cite{Zheng2012Physical, Wang2013Secure}.

Noteworthily, several existing beamforming techniques which are simple but not optimal have been also adopted widely in different scenarios of secure communications, such as null-space beamforming (also named zero-forcing beamforming) \cite{Dong2010Improving, Wang2012Distributed, Zheng2012Physical, Wang2013Secure, Yang2013Cooperative} and maximum ratio transmission (MRT) beamforming \cite{Wang2013Secure, Zhang2013On, Zhu2014Secure, Dong2015Energy}. Null-space beamforming chooses the beamforming vector lying in the null space of the eavesdropper's channel vector. Then, the eavesdropper gets nothing in the transmission process, such that the information leakage is avoided. In optimization designs, nulling signal at the eavesdropper can be expressed as a constraint, i.e.,
    \begin{equation}
    \label{null_space}
        \textbf{h}_{e} \textbf{w}^H = 0,
    \end{equation}
where $\textbf{h}_{e}$ and $\textbf{w}$ denote the eavesdropper's channel vector and the null-space beamforming vector, respectively. MRT is another attractive beamforming scheme because of its low computational complexity. MRT combined with maximal ratio combining (MRC) can maximize the signal-to-noise ratio (SNR) at the receiver and achieve a performance close to channel capacity in low-SNR scenarios \cite{Murthy2006Training}. In particular, the transmitter calculates its MRT beamforming vector, which only requires the knowledge of the channel from itself to the receiver. MRT beamforming can be expressed as
    \begin{equation}
    \label{MRT_beamforming}
        \textbf{w}  = \frac{\textbf{h}^H}{\| \textbf{h} \|},
    \end{equation}
where $\textbf{w}$ and $\textbf{h}$ denote the MRT beamforming vector and the legitimate channel vector from the transmitter to the intended receiver, respectively.

Precoding is another important technology to achieve different design objectives in physical layer security, which is especially appropriate for multi-stream data transmission or multi-user access. When the intended transceivers are equipped with multiple antennas, the confidential messages of one or multiple users can be spatially multiplexed onto multiple independent subchannels via precoding. By optimizing the precoder, the interested performance metrics of physical layer security can be achieved while the quality of service (QoS) can be guaranteed simultaneously \cite{Fakoorian2013On, Long2013Secrecy, Hanif2014On, Lin2014Secure, Wu2016Transmitter,Zhang2014Energy}. The secret dirty-paper coding (S-DPC) has been proposed to achieve the maximum secrecy rate in \cite{Liu2009Secrecy, Liu2010Multiple}. However, the complexity of S-DPC is computationally prohibitive, so that it is difficult to apply this precoding scheme in practice. The complexity of a precoding scheme may be crucial, which affects the application of precoding schemes in practice. In literature, due to the high complexity of the optimal precoding in some scenarios, as alternatives the suboptimal schemes have been developed to reduce computational complexity and facilitate their practical application \cite{Fakoorian2013On, Long2013Secrecy, Lin2014Secure, Wu2016Transmitter}. As a matter of fact, the linear precoding techniques are also attractive alternatives because of their simplicity \cite{Fakoorian2011Dirty, Fakoorian2013On}. As more simple linear precoding techniques, generalized singular value decomposition (GSVD) \cite{Fakoorian2011Dirty, Khisti2010SecureI, Khisti2010SecureII, Fakoorian2011MIMO, Huang2011Cooperative, Fakoorian2012Optimal} and regularized channel inversion (RCI) \cite{Geraci2012Secrecy, Geraci2013Physical, Geraci2013Large, Yang2014Confidential, He2015Base} have been extensively adopted in physical-layer secure transmission. The GSVD is to simultaneously diagonalize the legitimate channels and the wiretap channels, such that a set of parallel independent subchannels is created to transmit the messages of different users \cite{Fakoorian2011Dirty, Li2013Transmit}. Channel inversion (CI) precoding, sometimes known as zero-forcing (ZF) precoding, is a popular and practical linear precoding scheme which can control inter-user interference by canceling all signals leaked to the unintended users. RCI based on CI has better performance than plain CI by using a regularization parameter. RCI can achieve a tradeoff among signal power, interference, and information leakage \cite{Geraci2012Secrecy}.

Using AN to deteriorate the quality of the received signals at eavesdroppers is also a good way in physical layer security, which is herein referred to as noise-assisted secure strategies. In such strategies, the transmitted signal is superimposed with AN \cite{Liao2011QoS, Lin2012On, Romero2012Outage, Li2013Transmit, Zhang2013On, Poor2014Power, Zheng2015Optimal, Li2015RobustCoop, Wang2015Artificial, Wang2015Secrecy, HMWang2015Secure, Yang2015Optimal, Wang2013Joint, Goel2008Guar}. This strategy is also termed as ``masked beamforming'' in the multi-input single-output multi-eavesdropper (MISOME) wiretap channel \cite{Khisti2010SecureI}, ``masked precoding'' in the multi-input multi-output multi-eavesdropper (MIMOME) wiretap channel in \cite{Khisti2010SecureII}, and ``AN precoding'' in \cite{Poor2014Power}. In order to avoid interfering destination node, the simple but not optimal method is to let the AN lie in the null space of the signal space, i.e., satisfying
    \begin{equation}
    \label{null_space}
        \textbf{hz}=0,
    \end{equation}
where $\textbf{h}$ and $\textbf{z}$ denote legitimate channel vector and AN vector, respectively \cite{Poor2014Power, Wang2015Artificial, Wang2015Secrecy, HMWang2015Secure}. Furthermore, the AN can also be optimized globally to achieve the optimal secure performance \cite{Li2013Transmit, Zhang2013On, Li2015RobustCoop}.

\subsection{Antenna/Node Selection and Cooperation}

In multi-antenna and multi-node wireless networks, antenna and node selection have been exploited to strengthen transmission reliability, which also have great potential to enhance wireless security \cite{Zou2014Improving}. It has been verified that selecting the proper antennas or nodes from the candidate set is a simple but effective way to improve the performance of secure transmission while saving resource. As a result, antenna/node selection and cooperation have been considered and widely investigated in many works.

MIMO technologies are believed to be one of the foremost technologies pertaining to physical layer security. In a MIMO system, transmit antenna selection provides solutions to reduce the hardware complexity resulted from large antenna arrays and radio frequency chains, insertion losses attributed to radio frequency switches, and feedback overhead needed for transceiver communication \cite{Hanif2013Efficient}. In physical layer security, transmit antenna selection as a usual approach to exploit spatial degrees of freedom in multi-antenna scenarios, has been comprehensively investigated for maximizing the secrecy rate \cite{Hanif2013Efficient}, improving the SNR of the legitimate channels \cite{Yang2013Trans_antenna, Yan2014Transmit}, and enhancing security from the viewpoint of secrecy outage performance \cite{Alves2012Performance, Yang2013MIMO, Yang2013Physical, Wang2014Nakagami, Hu2015Secrecy}.

In multi-user networks, the randomness of users' geographical locations leading to random signal attenuation independently across users, can also be used to enhance secure performance \cite{Diggavi2004Great}. Accordingly, user selection/scheduling as a promising paradigm can be adopted to utilize the spatial diversity in multi-user networks \cite{Zou2014Improving, Kim2015Secure}. In a multi-user network, user selection determines which users should be scheduled for confidential data transmission. Typically, the user with the best channel quality is selected to improve secrecy rate or throughput \cite{Zou2013Physical, Zou2014Secrecy, Christopoulos2015Multicast}. This optimal selection scheme depends on both the legitimate and wiretap channels. Some suboptimal user selection schemes with considering wiretap links are also used due to their low complexity or the unavailability of wiretap CSI \cite{Zou2013Physical, Zou2014Improving, Zou2014Secrecy, Hoang2015Cooperative}. In addition, user selection/scheduling can also be used for saving power with secrecy rate constraints or enabling the largest possible user set with an effective transmission power constraint \cite{Mukherjee2009User}. In some situations, the legitimate channels to users may experience severe propagation loss and deep fading, and such users may have little chance to be scheduled. Therefore, the fairness of user selection/scheduling needs to be considered. Two competing problems should be balanced herein: achieving the optimal secure QoS while ensuring each user with certain opportunities to access networks \cite{Zou2014Improving, Pei2014On}.

In multi-relay cooperative networks, the distributed relay nodes may provide spatial degrees of freedom which can be exploited to improve secure QoS against the eavesdropping attack. It is well-known that cooperative relaying with relay selection can bring some benefits in terms of rate, EE, and security. More specifically, cooperative relaying combined with relay selection has the potential of maximizing the secrecy capacity \cite{Zou2013Optimal}, maximizing the Shannon capacity to the destination node as well as minimizing that to the eavesdropper \cite{Zou2013Optimal, Bao2013Relay}, reducing the secrecy outage probability \cite{Al2015Opportunistic, Wang2015Generalized}, maximizing the SNR ratio of the destination node to the eavesdropper \cite{Bao2013Relay, Fan2014Secure}, or saving the limited power of network nodes \cite{Nomikos2015Relay, Dong2015Energy, Dong2016Achieving}. Generally speaking, to strengthen the network security against the eavesdropping attack, three relay selection schemes have been proposed, which are referred to as minimum selection considering only the relay-eavesdropper links, conventional selection considering only the relay-destination links, and the optimal selection taking the both links into account \cite{Bao2013Relay, Fan2014Secure}. In literature, some heuristic algorithms have also been proposed for the optimal relay selection with different purposes.

Relay nodes can be used for not only cooperative relaying but also cooperative jamming \cite{Tekin2008The, Mukh2014Principles, Yener2015Wireless}. Cooperative jamming with jammer selection also has the ability to enhance secrecy of wireless networks. This security-enhanced strategy selects the jammers from trusted  or untrusted intermediate nodes to confuse eavesdroppers by transmitting artificial interference signals \cite{Yang2013Optimal, Lee2014Multiuser, Wang2015Uncoordinated, c2016Opportunistic}. With regard to the untrusted nodes which may be potential eavesdroppers, we should use them discreetly. However, it has been verified in \cite{He2010Cooperation} that, seeking for cooperation with the untrusted relay nodes can achieve a higher secrecy rate than just treating them as pure eavesdroppers. In other words, the untrusted relays can also be used for cooperative relaying while protecting the confidential data from them \cite{He2008The, Jeong2012Joint, Zhang2012Physical, Huang2013Joint, Mukherjee2013Imbalanced, Khodakarami2013Link, Huang2013Secure, Mo2014Secure, Richter2015Weak,Deng2015Secrecy}. Therefore, no matter whether the relays are trusted or not, they can be used intelligently for cooperative relaying or jamming \cite{He2008Two, Deng2015Secrecy}. Moreover, cooperative jamming with the destination node can also provide secrecy improvements, such as in \cite{Sun2012Performance, Huang2013Secure, Park2013On}.

\begin{figure}[t]
\centering
\includegraphics[width=2.5in]{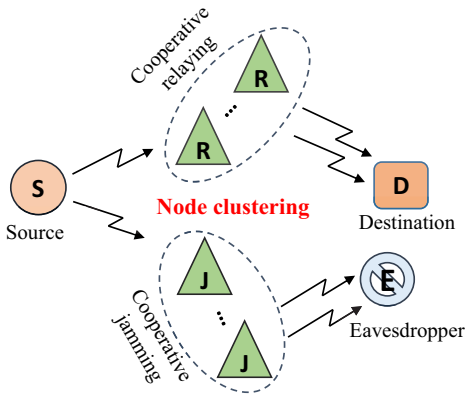}
\caption{An illustration of joint relay and jammer selection and cooperation.}
\label{Fig_relay_jammer_selection}
\end{figure}

In practice, the joint relay and jammer selection is more effective for improving the secure performance of the whole network than using any single approach. As illustrated in Fig. \ref{Fig_relay_jammer_selection}, by using such a joint method, some proper intermediate nodes are  selected to operate at a conventional relaying mode for assisting the confidential data transmission between the source node and the destination node. Meanwhile, another set of intermediate nodes are selected as jammers to confuse the potential eavesdropper \cite{Krikidis2009Relay, Chen2011Joint, Wu2013Effect, Liu2015Relay, Hui2015Secure, Wang2015Joint_RJ, Zhang2015Partner}.

\subsection{Joint Strategies of Several Approaches}

The secrecy improvements in physical layer can be supported by secure resource allocation, signal processing, and antenna/node selection and cooperation. Secure resource allocation mainly focuses on resource usage policies by fully using the multidimensional wireless resources involving frequency, timeslot and power. Secure beamforming and precoding belonging to signal processing are to design beamformer and precoder to well exploit the characteristics of multi-antenna and multi-node settings which may form MIMO or virtual MIMO networks. Antenna/node selection and cooperation aim at selecting the proper antennas or nodes from the candidate set to improve the performance of secure transmission. All of the foregoing strategies can be carried out to strengthen information security while achieving the requirements of performance metrics and resource savings. In other words, based on the ideas of these fundamental security strategies, we can design some different transmission schemes to achieve the specific requirements of network performance subject to secrecy and resource constraints.

In order to achieve better performance of physical layer security, any single approach mentioned above might not be sufficient. Therefore, the joint strategies based on some of the above approaches may be preferable in practical applications. As in \cite{Ng2011secure}, \cite{Luo2011User}, and \cite{Zhu2014Cross}, joint resource allocation and user scheduling have been proposed to enhance physical layer security in OFDMA networks. Antenna selection combined with beamforming/precoding has been demonstrated to be effective in secure MIMO system designs \cite{Hanif2013Efficient, Yan2014Transmit, Hanif2016Antenna}. Distributed beamforming with relay/jammer selection has been exploited in cooperative networks \cite{Kim2012Combined, Wang2013Hybrid, Wang2015Hybrid}. Additionally, other joint strategies have also been addressed for some specific scenarios to obtain secrecy improvements, such as cooperative beamforming and user selection in \cite{Krikidis2013Secrecy, Hoang2015Cooperative}, jamming-aided beamforming/precoding in \cite{Liao2011QoS, Lin2012On, Wang2013Joint, Liu2013Destination, Zhu2014Secure, Yang2014Joint, Wang2015JointCoop}, joint power allocation and beamforming/precoding in \cite{Fakoorian2012Optimal, Kha2013Joint, Wang2014Joint, Wang2015Joint}, etc.


\section{Performance metrics and basic optimization problems in physical layer security} \label{section_metrics}


For the secure transmission designs, the choice of performance metrics is remarkably critical. In physical layer security designs, there are several problems usually being raised from different performance requirements:
\begin{enumerate}
  \item The transmission effectiveness of secure transmission strategies, that is evaluated by the achievable secrecy rate/capacity.
  \item The reliability of secure transmission strategies, which is measured by secrecy outage probability/capacity.
  \item The power cost of secure transmission strategies, that is the minimum power consumption needed for ensuring the secure QoS.
  \item The EE of secure transmission strategies, which focuses on the amount of secret bits transferred with unit energy or the energy consumption required for sending one secret bit.

\end{enumerate}

To investigate these problems listed above, the corresponding metrics termed as secrecy rate/capacity, secrecy outage probability/capacity, power consumption, and secure EE, are usually adopted in system designs to evaluate the achievable performance of the proposed secure transmission strategies. More specifically, these performance metrics are usually taken as the optimization objectives for system designs in different application scenarios.


\subsection{Secrecy Rate/Capacity}

Being similar to the data rate in conventional communications, the secrecy rate is a fundamental metric to assess the transmission effectiveness of physical-layer secure strategies. In physical layer security, the secrecy rate is defined as the secret bits transmitted on the given channel per second, which heavily depends on channel inputs. To evaluate the secrecy more conveniently and computation affordably, the Gaussian inputs as well as the achievable secrecy rate are usually adopted in many works \cite{Wang2015Enhancing}. The achievable secrecy rate can be described as the difference between the achievable data rate of the legitimate channel and the wiretap channel with the Gaussian codebook, which is expressed as
    \begin{equation}
    \label{Rs}
        R_s = \left[R_m  - R_e\right]^+,
    \end{equation}
where $[x ]^+ \triangleq \max\{0,x\}$. $R_m$ denotes the data rate of the legitimate channel from the source node to the destination node. $R_e$ denotes the data rate of the wiretap channel from the source node to the eavesdropper. Clearly, the achievable secrecy rate is a lower bound of the secrecy capacity\cite{Wang2015Enhancing}. In practical designs, by some approaches as secure beamforming or resource allocation, a non-zero secrecy rate can be obtained since the wiretap channel is intentionally degraded while improving the quality of legitimate channel.

Another metric closely related to secrecy rate is secrecy capacity, which is defined as the upper bound of the secrecy rate \cite{Liang2010Information} \cite{bloch2011Security}. More specifically, the secrecy capacity is the maximum secrecy rate by which the confidential messages of the source node can be securely and reliably transmitted to the destination node whereas the unauthenticated users cannot obtain any useful information in this process. In Wyner's pioneering work \cite{Wyner1975wiretap}, the secrecy capacity of a degraded wiretap channel has been given by
    \begin{equation}
    \label{Cs}
        C_s = \sup\limits_{p(X)} \{ I(X; Y) - I(X; Z) \},
    \end{equation}
where $X$ denotes the channel inputs at source node. $Y$ and $Z$ denote the channel outputs at the destination node and eavesdropper, respectively. $I(\cdot; \cdot)$ represents the mutual information. The secrecy capacity shown in \eqref{Cs} can be achieved by choosing the optimal input probability distribution $p(X)$. For any distribution $p(X)$, the corresponding $X$, $Y$, and $Z$ form a Markov chain \cite{Liang2010Information} \cite{bloch2011Security}.

Based on Wyner's results, Csisz$\acute{\text a}$r and K$\ddot{\text o}$rner investigated a more general (non-degraded) wiretap channel and derived its secrecy capacity as \cite{Csiszar1978Broadcast}

    \begin{equation}
    \label{Cs_general}
        C_s = \sup\limits_{p(V, X)} \{ I(V; Y) - I(V; Z) \},       
    \end{equation}
where $V$ is an auxiliary input variable. By introducing appropriate random variable $V$, the maximization in \eqref{Cs_general} can be implemented over all joint probability distributions $p(V, X)$ while forming a Markov chain $V \rightarrow X \rightarrow (Y,Z)$. As to the familiar Gaussian channel, the secrecy capacity has been derived in \cite{Leung1978Gaussian} as following:
    \begin{equation}
    \label{Cs_gaussian}
        C_s = C_m - C_e ,
    \end{equation}
where $C_m$ and $C_e$  denote the Shannon capacities of the legitimate and wiretap channels, respectively.

The aforementioned secrecy rate and secrecy capacity have been investigated without considering the fading of wireless channels. However, fading of wireless channels is an inevitable issue in many situations, as stated in \cite{bloch2011Security} in which three standard fading models as well as the corresponding ergodic secrecy rate/capacity have been well discussed, including the ergodic-fading model, block-fading model, and quasi-static fading model. When the channel fading is taken into consideration, the average capability of secure communication over fading channels should be evaluated, and the ergodic secrecy rate or secrecy capacity is then a quite suitable metric for this case \cite{Yingbin2008Secure, Wang2015Enhancing}. In practice, since achieving ergodic secrecy capacity may be computationally difficult in many situations, the achievable ergodic secrecy rate is therefore adopted to measure the secrecy performance in fading scenarios. The achievable ergodic secrecy rate is defined as the difference between the ergodic rates of the legitimate and wiretap channels with Gaussian codebooks, which is more computationally efficient in many cases. As the lower bound of ergodic secrecy capacity, the achievable ergodic secrecy rate has usually been taken as the optimization objective in secure transmission designs with the consideration of channel fading \cite{Wang2015Enhancing}.

Towards secure communication system designs, the primary concern is how much the secrecy rate can be achieved for delivering the confidential data securely and reliably. This problem can be modeled as maximization of the achievable secrecy rate, that is to maximize the achievable secrecy rate as much as possible by using some physical-layer technologies such as resource allocation, beamforming/precoding, cooperative diversity, or other optimization algorithms. To maximize the achievable secrecy rate, the most important factor is the power limitation in addition to the bandwidth. Accordingly, one common formulation of achievable secrecy rate maximization on the given channels generally aims at maximizing the secrecy rate under the constraints of the maximum allowed power. For instance, the achievable secrecy rate maximization in a relay network can be modeled as
    \begin{equation}
    \label{Rs_max}
    \begin{array}{l}
        \mathop{\max\limits_{P_t^{(S)}, P_t^{(j)}, j \in \Omega} R_s \left( P_t^{(S)}, P_t^{(j)} \right)} \\
        \text{s.t.}~~
            P_t^{(S)}  + \sum\limits_{j \in \Omega} P_t^{(j)} \le P_{max}^{sum},  \\
        ~~~~~~\text{or}
                \left\{\begin{array}{l}
                   0 \le P_t^{(S)}  \le  P_{max}^{(S)},  \\
                   0 \le P_t^{(j)}  \le  P_{max}^{(j)}, j \in \Omega,
                \end{array}\right.
    \end{array}
    \end{equation}
where $P_{sum}^{max}$ denotes the maximum sum transmission power of all nodes in the relay network, and $P_{max}^{(S)}$ and $P_{max}^{(j)}$ denote the maximum transmission power of the source and the $j$th relay nodes, respectively. In existing literature, there are two kinds of power constraints in the problem of secrecy rate maximization. One is the sum power constraints of all nodes specified by the constraint $P_t^{(S)}  + \sum\limits_{j \in \Omega} P_t^{(j)} \le P_{max}^{sum}$ in \eqref{Rs_max}, and the other is the individual power constraint of each node specified by the constraints $0 \le P_t^{(S)}  \le  P_{max}^{(S)}$ and $0 \le P_t^{(j)}  \le  P_{max}^{(j)}, j \in \Omega$ in \eqref{Rs_max}. Noteworthily, beamforming may be more effective for maximizing secrecy rate by strengthening signals on a desired direction and suppressing/eliminating signals on undesired directions. When beamforming is considered in such a relay network, the weight vector of all relays will be introduced to replace power, as investigated in \cite{Dong2010Improving} and \cite{Li2011OnCoop}. The maximization of the achievable secrecy rate has been comprehensively investigated in many scenarios, for example multicarrier systems, multi-antenna systems, multi-node cooperative systems, etc.

\subsection{Secrecy Outage Probability/Capacity}

Due to the channel fading and imperfect CSI, secure transmission may be broken. Therefore, it is of particular interest to explore the secrecy outage behaviour of a secure transmission strategy \cite{Bloch2008Wireless, Yingbin2008Secure}. Then, the secrecy outage probability is an appropriate metric to characterize the probability that secure transmission cannot be achieved. Precisely, the secrecy outage probability is defined as the probability that a secrecy outage event happens.

There are two different definitions of secrecy outage events. The more popular one is that the secrecy outage happens when the instantaneous secrecy capacity $C_s$ drops below a target secrecy rate $R_s^0$, i.e., $\{C_s < R_s^0\}$ \cite{Gerbracht2012Secrecy, Barros2006Secrecy, Bloch2008Wireless, Bashar2011secrecy, Park2013On, Xiong2012A}. In other words, the target secrecy rate is too high to be supported by the current channel state, and the information security is compromised. The secrecy outage probability of this definition is given by
    \begin{equation}
    \label{p_out1}
        p_{out}\left( R_s^0 \right) = \text{Pr}\left\{C_s < R_s^0\right\}.
    \end{equation}
In \eqref{p_out1}, the outage events $\{C_s < R_s^0\}$ happen whenever the intended receiver does not receive the secret messages reliably (i.e., the message cannot be decoded correctly by intended receivers) or the message transmission is not perfectly secure (i.e., some information may leak to eavesdroppers) \cite{Yuksel2011Diversity, Zhou2013Rethinking}. However, this definition does not distinguish between reliability and security. As a result, an outage based on this definition does not necessarily imply a failure in achieving perfect secrecy. To be specific, the outage events $\{C_m < R_s^0\}$ mean that the secrecy rate cannot be supported by the legitimate channels and the secure transmission would be certainly suspended. Clearly, these suspension events fall within the outage events $\left\{C_s < R_s^0\right\}$ due to $C_m < R_s^0$ implying $C_s < R_s^0$, but it is clearly not a failure in achieving perfect secrecy \cite{Zhou2013Rethinking}. Then, the outage probability in secure transmissions can be more explicitly expressed as \cite{Bo2014Outage}
    \begin{equation}
    \label{p_out2}
        p_{out} \left( R_m^0, R_s^0 \right) = 1- \text{Pr}\left\{C_m \ge R_m^0 , C_s \ge R_s^0\right\},
    \end{equation}
where $R_m^0$ is the target coding rate of the confidential message, $R_s^0 \le R_m^0$.  The outage events $\{C_m < R_m^0\}$ imply the legitimate channels cannot support the coding rate $R_m^0$. Consequently, at a target secrecy rate $R_s^0$ and a target coding rate $R_m^0$, the reliable and secure transmission can only be ensured at a probability $1-p_{out}\left( R_m^0, R_s^0 \right)$.

The other definition of secrecy outage is proposed in \cite{Zhou2013Rethinking}, which directly measures the probability that a transmitted message fails to achieve perfect secrecy. In \cite{Zhou2013Rethinking}, considering the Wyner's encoding scheme \cite{Wyner1975wiretap}, the rate difference $R_e \triangleq R_m - R_s $ is defined to reflect the cost of securing message transmission against eavesdropping, where $R_m$ and $R_s$, respectively, denote two rates chosen by secure encoder, namely, the rate of the transmitted codewords and the rate of the confidential information. The transmitted messages can be decoded correctly if $R_m<C_m$, whereas it fails to achieve perfect secrecy if $R_e < C_e$. Hence, the secrecy outage probability is defined in \cite{Zhou2013Rethinking} as
    \begin{equation}
    \label{p_out3}
        p_{out} \left( C_e \right) \!=\! \text{Pr}\left\{R_m \!-\! R_s \!<\! C_e  |  \text{message transmission} \right\},
    \end{equation}
which is a probability conditioned upon a message actually being transmitted. If the source transmitter has no knowledge about the instantaneous CSI of the legitimate channel, the transmission may always occur, so that the secrecy outage probability in \eqref{p_out3} then reduces to the unconditional probability $\text{Pr}\left\{R_m - R_s < C_e \right\}$. More generally, when the instantaneous CSI of legitimate channel is available, the source transmitter can decide whether or not to transmit with possibly variable rates according to channel conditions. Therefore, it is possible to reduce the secrecy outage probability by carefully designing the rate of the transmitted codewords $R_m$, the rate of the confidential information $R_s$, and the condition for transmission \cite{Zhou2013Rethinking}.

Another important concept related to the secrecy outage probability is the secrecy outage capacity $C_{out}(\epsilon)$, which is defined as the largest secrecy rate that can be supported under a tolerable secrecy outage probability $\epsilon$ \cite{Wang2015Enhancing} \cite{Barros2006Secrecy} \cite{Gungor2013Secrecy} \cite{Chen2014Energy}. In other words, the secrecy outage capacity is the maximum achievable secrecy rate such that the secrecy outage probability is less than $\epsilon$, i.e.,
    \begin{equation}
    \label{p_out_C}
        p_{out} \left( C_{out}(\epsilon) \right) = \text{Pr}\left\{ C_s <  C_{out}(\epsilon) \right\} = \epsilon.
    \end{equation}

The practical significance of secrecy outage probability/capacity is that these definitions provide outage formulations which give a more explicit measure of the security level. From the system design perspective, it is meaningful to evaluate the secrecy outage behaviour of the proposed transmission scheme \cite{Zhou2013Rethinking}.


For the optimization design in physical layer security, the reliability of secure transmission which is generally measured by secrecy outage probability has also attracted increasing concerns. Ideally, the secure communication should be implemented without outage. Motivated by this observation, we expect to reduce the secrecy outage probability with the best effort. This raises the optimization problem of secrecy outage probability minimization subject to resource constraints. Taking the relay network as an example, the minimization of the secrecy outage probability can be roughly formulated as
    \begin{equation}
    \label{Outage_min}
    \begin{array}{l}
        \mathop{\min \limits_{P_t^{(S)}, P_t^{(j)}, j \in \Omega} ~p_{out}\left( R_s^0 \right) } \\
        \text{s.t.}~~
        \left\{\begin{array}{l}
              0 \le P_t^{(S)}  \le  P_{max}^{(S)},  \\
              0 \le P_t^{(j)}  \le  P_{max}^{(j)}, j \in \Omega.
        \end{array}\right.
    \end{array}
    \end{equation}
In \eqref{Outage_min}, the peak power of each transmission node is taken into account to limit the excessive high power consumption resulted from the improvement of the secrecy rate in minimizing secrecy outage probability.

\subsection{Power/Energy Consumption}

Power/energy consumption is a key consideration in resource-limited scenarios such as battery-dependent networks. In general, the sustainability of secure communications in such networks is the most important concern. Therefore, to reduce energy consumption and prolong network lifetime, power/energy cost is one primary metric considered in physical layer security designs.

Before designing a secure transmission scheme with limited power and energy, we first analyse the factors of power consumption in wireless networks \cite{Chen2014Energy}. According to \cite{Cui2004Energy}, the total power consumption along the signal path can be divided into two main components: the power consumption of all the power amplifiers $P_{a}$ and the power consumption of all other circuit blocks $P_c$. The power consumption of all power amplifiers heavily depends on the output power of power amplifiers $P_t$, i.e.,
    \begin{equation}
    \label{PA}
        P_a = P_{t} / \eta,
    \end{equation}
where $\eta$ is the efficiency of power amplifier. The other circuit blocks include the basic circuits at the transmitter and receiver excluding power amplifiers, such as active filter, frequency synthesizer, mixer, intermediate frequency amplifier, analog-to-digital or digital-to-analog converter, and so on  \cite{Chen2014Energy, Cui2004Energy}. Accordingly, the power consumption of all other circuit blocks $P_c$ can be roughly expressed as \cite{Cui2004Energy, Chenzi2012}
    \begin{equation}
    \label{PC}
        P_c = N_t P_{ct} + N_r  P_{cr} + P_{c0},
    \end{equation}
where $N_t$ and $N_r$ denote the numbers of transmitter antennas and receiver antennas, respectively. $P_{ct}$ and $P_{cr}$ denote the power consumed by the basic circuits at each transmit and receive chain, respectively. $P_{c0}$ denotes the power consumed by baseband circuits such as digital signal processing circuits. It can be seen that $P_{ct}$, $P_{cr}$, and $P_{c0}$ are independent of the secrecy rate. As a result, the total power consumption of a system can be given by
    \begin{equation}
    \label{P_tot}
    \begin{aligned}
        P &= P_{a} + P_c \\
          &= P_{t} / \eta + N_t P_{ct} + N_r  P_{cr} + P_{c0}.
    \end{aligned}
    \end{equation}

The power consumption of a wireless communication system can be usually formulated as \eqref{P_tot}. However, in a practical scenario, there may be some variations in the power consumption model. For example, in a cooperative relay network, the power consumption model can be expressed as
    \begin{equation}
    \label{power_relay}
        P = \frac{1}{2\eta }\left( P_t^{(S)} + \sum\limits_{j \in \Omega} P_{t}^{(j)} \right) + P_c^{(S)} + \sum\limits_{j \in \Omega} P_c^{(j)},
    \end{equation}
where, $\Omega$ is the set composed of relay nodes and $j$ denotes the $j$th relay node. $P_t^{(S)}$ and $P_{t}^{(j)}$ denote the transmission power of the source node and the $j$th relay node, respectively. $P_c^{(S)}$ and $ P_c^{(j)}$, respectively, denote the power of the basic circuit blocks at the source node and the $j$th relay node, which can be obtained by \eqref{PC}. The factor $\frac{1}{2}$ lies in the fact that the transmission is completed in two stages due to half duplex.

The resource-limited regime motivates us to develop the power-efficient transmission strategies which aim at minimizing power consumption \cite{Marques2008Power,Joung2012Power,Nomikos2015Relay}. For this purpose, the power level of transmitters should be adjusted to save transmission power while satisfying the target QoS requirements. It is worth noting that, although relay cooperation has the potential of transmission effectiveness, reliability, and security, relay nodes may consume additional power, such as the basic circuit power which is inherent in relay cooperation and unrelated with secrecy rate. Therefore, from the viewpoint of transmission designs, the power adaptation and relay selection should be performed jointly to achieve the requirements of  power-efficient secure transmission.

It is noteworthy that a higher transmission rate of messages can be achieved if no secrecy constraint is imposed. When secrecy is considered, the transmission rate of confidential messages will decrease due to secure coding. Hence, higher power consumption is needed to ensure a higher level of secrecy at the physical layer \cite{Comaniciu2013An}.

For secure transmission designs in power-limited scenarios, such as the transmission nodes powered by batteries or energy harvesting devices \cite{Ozel2011Trans}, we should give priority to saving power and prolonging communication durations. These observations motivate us to design secure transmission schemes focusing on the minimization problem of power cost. In general, power minimization means consuming the minimum power to achieve the fundamental demand of secure transmission such as the minimum target secrecy rate \cite{Dong2010Improving, Li2011OnCoop, Liu2014Transmit, Nomikos2015Relay}, the required SNR threshold of destination node \cite{Wang2013Joint, Lei2011Secure}, the given probability of secrecy \cite{Romero2012Outage}, or other performance requirements. For example, in a relay network, the basic formulation of power minimization can be expressed as
    \begin{equation}
    \label{power_min}
    \begin{array}{l}
        \mathop{\min \limits_{P_t^{(S)}, P_t^{(j)}, j \in \Omega} \! P \!=\! \frac{1}{2\eta } \! \left( \! P_t^{(S)} \!+\! \sum\limits_{j \in \Omega} \! P_{t}^{(j)} \right) \!+\! P_c^{(S)} \!+\! \sum\limits_{j \in \Omega} \! P_c^{(j)} \! } \\
        \text{s.t.}~~
           R_s \ge R_s^0.
    \end{array}
    \end{equation}
The formulation in \eqref{power_min} is only a rough model, which can be specified in practical applications. For instance, when the beamforming is performed for minimizing the power consumption, the total power is then determined by the weights of beamformer \cite{Dong2010Improving, Li2011OnCoop, Li2014Power}.

\subsection{Secure EE}

In the conventional communications without secrecy constraints, the utilized efficiency of system energy referred to as EE is an important metric for green transmission strategy designs. When the security threats and energy limitations are considered jointly in wireless networks, it is significant to design energy-efficient secure transmission strategies which should operate in a confidential and green manner. Therefore, from the perspective of green physical layer security, an appropriate metric for assessing the utilized efficiency of system energy is also of primary importance. In general, the utilized efficiency of system energy can be measured by different metrics from different viewpoints, such as the viewpoints from the component level, equipment level, and system/network level. Towards the EE of system/network level, it aims at measuring both the energy consumed by all communication nodes and the performance experienced at the network level (i.e., capacity, security, coverage, etc.). The EE of system/network level is popular in transmission strategy designs.

There are two main metrics which have been defined for evaluating the EE of novel techniques towards physical layer security. One metric is the secure EE \cite{Dong2015Energy, Dong2016Achieving, Dong2016Green}, which is defined as the amount of secret bits transmitted with unit energy consumption. Designing energy-efficient secure transmission strategies with this metric, it is expected to maximize the secure EE. The resulting effect is that as much confidential information as possible is transmitted with a given amount of energy. Hence, given the amount of energy $\Delta E$ consumed in a duration $\Delta T$, the secure EE can be defined as
    \begin{equation}
    \label{EE_bits_J}
        E_B = \frac{R_s \Delta T}{\Delta E} = \frac{R_s}{P} \left( \text{bits} / \text{Joule} \right).
    \end{equation}
The metric of secure EE is in fact the ratio of secrecy rate to total power consumption, which has been frequently used in literature for investigating the EE of physical-layer secure communications \cite{Chen2013Energy, Zhang2014Energy, Dong2015Energy, Dong2016Achieving, Dong2016Green}. This metric is also termed as ``secret bits per Joule'', since its unit is bits/Joule.

Another metric proposed to assess the EE of physical-layer secure transmissions is the energy per secret bit, which is suitable for evaluating the minimum energy required to send one secret bit (i.e., minimum bit energy required for reliable communications under secrecy constraints). The precise formula of this metric is the ratio of total power consumption to secrecy rate \cite{Gursoy2012Secure} \cite{Comaniciu2013On}, i.e.,
    \begin{equation}
    \label{EE_joule_bit}
        E_J = \frac{P}{R_s} \left( \text{Joules} / \text{bit} \right).
    \end{equation}

Noteworthily, these two metrics of secure EE are reciprocal to each other. The resulting optimization problem by using one metric is in general the dual problem of that by using the other metric. Which metric is better in practice should fully consider the practical scenarios, for reducing the difficulties of secure transmission designs. As stated in \cite{Mahapatra2016}, the metric of secret bits per Joule is more popular since it is convenient to capture the degree of proportionality between the energy consumption and different levels of load. This metric can reflect dynamic network conditions considering energy consumption and secrecy constraints in different situations of load. However, the metric of energy per secret bit is suitable to assess the network EE only at a nonzero secrecy rate.

In addition, it is obvious that the metrics of secure EE are closely related to the model of the power consumption. The traditional energy-efficient technologies only consider the transmission power, but which is not the only part of power consumption in a networks. A holistic and system-wide power model is imperative \cite{Feng2013A}. Therefore, the secure EE should be formulated with all power consumption including transmission power, basic circuit power, and signaling overhead in the entire network \cite{Feng2013A}.

In physical layer security, more power and energy, compared with the conventional communication without secrecy requirement, may be consumed to protect confidential information against eavesdropping. This observation can be verified by the secrecy rate function shown in \eqref{Rs} where the information rate leaking to the eavesdropper generates extra consumption of power and energy. This fact may increase the burden of power and energy supplies, in particular in the scenarios with limited power and energy. When the limited power and energy become the main factors for securing communications, the first concern, impelled by the requirements of ``green communication'', is to deliver confidential information with high secure EE as much as possible. This motivation raises the maximization of the secure EE in physical layer security. Also taking the relay network as an instance, the mathematical formulation of secure EE maximization can be roughly modeled as
    \begin{equation}
    \label{secure_EE_max}
    \begin{array}{l}
        \mathop{\max \limits_{P_t^{(S)}, P_t^{(j)}, j \in \Omega}~\left\{E_B = \frac{R_s}{P}\right\} }\\
        \text{s.t.}~~
        \left\{\begin{array}{l}
               0 \le P_t^{(S)}  \le  P_{max}^{(S)}, \\
               0 \le P_t^{(j)}  \le  P_{max}^{(j)}, j \in \Omega,  \\
               R_s \ge R_s^0.
        \end{array}\right.
    \end{array}
    \end{equation}
It is worth noting that the secure EE maximization should ensure the secure QoS requirement which is specified by the constraint $R_s \ge R_s^0$ in \eqref{secure_EE_max}. Here, $R_s^0$ is used to avoid achieving high secure EE but with too low secrecy rate. In some literature \cite{Ng2012Energy, Ng2012EnergyOFDMA}, $R_s^0$ can be adjusted to balance the system performance between secure EE and secrecy rate.

It is pointed out that the aforementioned metrics and optimization problems are all based on information-theoretic security, since those metrics and problems are intertwined with the secrecy rate/capacity which is based on information theory. According to \cite{PHY2017SecuGAP}, another type of performance metrics for secrecy is based on practical measures where the secrecy level is quantified by the metrics that can be observed in practical communication scenarios, such as secrecy gap which is usually characterized by bit error rate or packet error rate versus SNR. To be specific, secrecy gap reflects the minimum required difference between the SNR of legitimate receiver and eavesdropper for which secure communication is possible \cite{PHY2017SecuGAP, Baldi2013Codgap}. This metric has also been used to make a quantitative measure for system designs, for instance in \cite{Hamamreh2017OFDM, Baldi2013Codgap, Hamamreh2017OSTBC}.

\section{The state of the art of optimization and design in physical layer security} \label{section_state_art}

In the previous section, we discussed the performance metrics and fundamental optimization problems in physical layer security. Each research topic of physical-layer security designs investigated in Section \ref{section_secure_topics} involves extending these fundamental optimization problems to practical scenarios according to specific application conditions and solving the resulting optimization problems to achieve the required performance metrics. In this section, the state of the art of optimization and design in physical layer security will be summarized from the perspectives of the aforementioned research topics in physical-layer security designs. Each research topic will be presented from four categories of fundamental optimization problems including maximization of achievable secrecy rate, minimization of secrecy outrage probability, minimization of power consumption, and maximization of secure EE.

\subsection{Secure Resource Allocation}

As a promising way for improving the performance requirements of physical layer security, secure resource allocation has been extensively investigated for different purposes. As discussed above, the designs of secure resource allocation are usually performed by solving four optimization problems which are related with the corresponding performance metrics.

\subsubsection{Maximization of achievable secrecy rate}
Many works focus on designing secure resource allocation strategies to improve achievable secrecy rate. A conventional approach towards maximizing secrecy rate in multicarrier systems is to globally allocate the limited power and subcarriers for all transmission nodes. This goal usually leads to a mixed integral programming in many scenarios, which has been investigated in many works \cite{Jindal2015Resource, Jeong2011Optimal, Tao2011Subcarrier, zhang2015commun, M2016Artificial, Karach2015Security}. Such as in \cite{Jindal2015Resource}, the resource allocation for a secure multicarrier AF relay communication system is investigated, in which decision variables $\mu_{si} \in \{ 0, 1\}$ and $\mu_{ri} \in \{ 0, 1\}$ are defined for the source and the relay, respectively, for specifying the state of communication on a carrier $i$. More specifically, if $\mu_{si}=1$ and $\mu_{ri}=1$ then both the source and the relay transmit in respective slots, while if $\mu_{si}=1$ and $\mu_{ri}=0$ then only the source transmits in its slot and it remains silent with the relay in the second slot. The case $\mu_{si}=0$ and $\mu_{ri}=0$ indicates no communication in both the slots and the case $\mu_{si}=0$ and $\mu_{ri}=1$ has no significance. Then, the resource allocation strategy for maximizing secrecy capacity in such a relay-aided multicarrier system can be derived by solving the typical mixed integral programming
    \begin{equation}
    \label{resource_power_carrier}
    \begin{array}{l}
        \mathop{\max \limits_{P_{si}, P_{ri}, \mu_{si}, \mu_{ri}}  \sum \limits_{i} C_i \left( P_{si}, P_{ri}, \mu_{si}, \mu_{ri} \right)}\\
        \text{s.t.}~~
        \left\{\begin{array}{l}
               \sum \limits_{i} \mu_{si} \left( P_{si} + \mu_{ri} P_{ri} \right) \le P_{sum}^{max}\\
               P_{si} \ge 0, P_{ri} \ge 0 \\
               \mu_{si} \in \{ 0, 1\}, \mu_{ri} \in \{ 0, 1\},
        \end{array}\right.
    \end{array}
    \end{equation}
where $P_{si}$, $P_{ri}$, and $C_i$ denote the source power, the relay power, and the secrecy capacity on carrier $i$, respectively.

Other specific formulations towards different scenarios have also been explored in this areas. In \cite{Tao2011Subcarrier}, a secure resource allocation policy is addressed for a downlink OFDMA-based network with the coexistence of secure users and normal users which have no confidential messages and do not care about security issues. In \cite{Jeong2011Optimal}, the transmission modes referred to as no communication, direct communication, and relay communication are determined adaptively by subcarrier allocation while the optimal source and relay power allocation policy over all subcarriers is addressed to maximize the sum secrecy rate under a total power constraint. Jamming and AN-aided resource allocation for sum secrecy rate maximization is, respectively, studied in \cite{zhang2015commun} and \cite{M2016Artificial}, where the former focuses on the OFDMA-based two-way relay wireless sensor networks while the latter focuses on the OFDMA systems with joint secrecy information and power transfer. For considering the fairness of resource allocation in secure multiuser OFDMA downlink works, the work presented in \cite{Karach2015Security} aims to assign subchannels and allocate power to optimize the max-min fairness criterion over the users' secrecy rate. Besides, robust secure resource allocation in relay-assisted cognitive radio networks is investigated in \cite{Mokari2015SecureRobust} considering the uncertainty of CSI.

\begin{figure}[t]
\centering
\includegraphics[width=3 in]{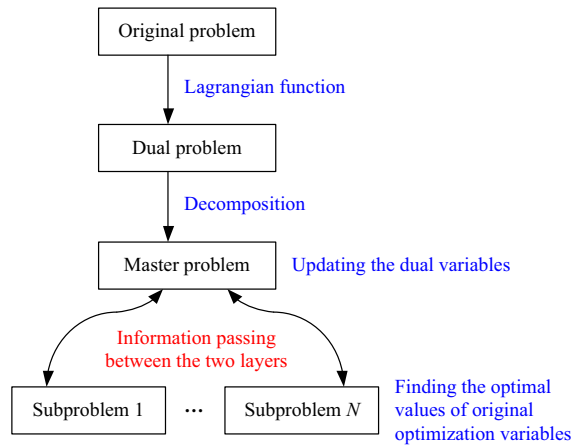}
\caption{Dual decomposition approach for secure resource allocation.}
\label{Fig_Decomposition}
\end{figure}

To solve the problems of secure resource allocation mentioned above, the approach of dual decomposition is usually adopted in many foregoing works. The basic idea of dual decomposition can be summarized as: 1) constructing a Lagrangian dual problem associated with the original problem by transforming the constraints into the objective function in the form of a weighted sum, and 2) decomposing the Lagrangian dual problem into distributed subproblems which are then coordinated with a high-level master problem by iterative alternating optimization between the two levels \cite{Palomar2006Decomp}, as illustrated in Fig. \ref{Fig_Decomposition}. Based on dual decomposition, the resource allocation in some secure scenarios can be solved by different distributed algorithms which are efficient for computing in many cases.

\subsubsection{Minimization of secrecy outage probability}

Secure resource allocation is also an effective approach for minimizing secrecy outage probability. Considering a typical secure OFDMA downlink system in \cite{Bo2014Outage}, the outage-optimal subcarrier allocation is addressed to minimize the secrecy outage probability $p_{out}^m $ of each user $m$ while guaranteeing that each user has the identical probability to access each subcarrier $n$. The formulation of the problem is summarized as
    \begin{equation}
    \label{Min_SOP}
    \begin{array}{l}
        \mathop{\min  \left\{ p_{out}^m \right\}_{\forall m }}\\
        \text{s.t.}~~
        \left\{\begin{array}{l}
               \sum \limits_{\forall m} \mu_{mn} \le 1  \\
               \sum \limits_{\forall n} \mu_{mn} \le 1  \\
               \mu_{mn} \in \{ 0, 1\},
        \end{array}\right.
    \end{array}
    \end{equation}
where $\mu_{mn}$ are the decision variables with $\mu_{mn} = 1$ meaning that subcarrier $n$ is assigned to user $m$. Otherwise, $\mu_{mn} = 0$. The constraints $\sum \limits_{\forall m} \mu_{mn} \le 1$ and $ \sum \limits_{\forall n} \mu_{mn} \le 1$ imply that each subcarrier can only be assigned to no more than one user with identical probability. It is noted that, to deal with the difficulty of such a probabilistic integral programming, a random bipartite graph approach is proposed with a logarithm-polynomial complexity when applying parallel implementations. A more complicated formulation of probabilistic mixed integral programming is investigated in \cite{Wang7801163min} to minimize the secrecy outage probability of a wireless systems with adaptive transmission rates and secrecy rates, in which a stochastic network optimization framework is introduced to overcome the difficulty of such a problem.

The outage-optimal power allocation is also explored extensively. By deriving explicit expressions of the secrecy outage probability, the closed-form solutions  of the optimal power allocation are obtained to achieve high outage performance in an AF relay network with destination-assisted jamming \cite{Park2013On}, an AN-aided secure multi-antenna transmission coexisting with randomly distributed eavesdroppers \cite{Zheng2015Optimal} \cite{Zheng2015LocBased}, and a MISO system with a multi-antenna eavesdropper \cite{Yang2015Optimal}, respectively.

The minimization of the secrecy outage probability is also raised in the scenarios of secure wireless information and power transfer in \cite{Xing2016Secrecy, He2014Onplacement, YU2016OuT}. In \cite{Xing2016Secrecy}, the transmission power allocation and power splitting ratio for AN signal are jointly
optimized to minimize the outage probability for delay-limited secrecy information transmission based on the approaches of dual decomposition and alternating optimization. In \cite{He2014Onplacement}, the minimum secrecy outage probability is achieved by optimizing the optimal placement of energy harvesting node with physical layer security considerations. In \cite{YU2016OuT}, the secrecy outage probability minimization problem and the average harvested energy maximization problem in wireless information and power transfer systems are solved by an optimization framework of target secrecy rate and power allocation ratio.

It is worth noting that the minimization of the secrecy outage probability is dual to the maximization of the secrecy outage capacity which is another optimization design related to secrecy outage performance. As in \cite{Ng2011secure}, the packet data rate, secrecy data rate, power, and subcarrier allocation policies of an OFDMA DF relay network are designed to maximize the average secrecy outage capacity by the dual decomposition and gradient method. In\cite{Chen2015Opti}, the solutions of the optimal relay power allocations for a massive MIMO DF relay network are derived for maximizing the secrecy outage capacity and minimizing the interception probability, respectively. The results in \cite{Chen2015Opti} are expanded in \cite{Chen2016Resource}, in which to cope with the nonconvexity of the joint node power and transmission time allocation problem, the approach of alternating optimization is addressed by maximizing over some of the variables and then maximizing over the rest.

\subsubsection{Minimization of power consumption}

\renewcommand\arraystretch{1.5}
\begin{table*}[t]
\centering
\caption{ The comparison of power-efficient resource allocation in different scenarios}
\label{tab_resource_power_min}
\begin{tabular}{  p{2.7 cm} p{2.5cm}  p{3.5cm}   p{2.7cm}   p{4 cm}   }

\hline

\hline
Scenarios  &  Wireless resources  &  Assumptions of CSI   &   Secure QoS constraints   &  Solutions  \\
\hline

\hline
  MISO networks with AN \cite{Romero2012Outage}  & The total  power for secrecy information and AN  &  Perfect CSI of legitimate channel and unknown CSI of eavesdropper   & Received SNR at destination and a given probability of secrecy  &     A closed-form solution \\
\hline
  Multiuser MISO networks with jamming \cite{Akgun2017Exploit}  & The total power for secrecy information and jamming signals &  Perfect CSI of all channels  &  Target secrecy rate  &  Numerical analyses based on a line search method \\
\hline
 A non-orthogonal multiple access system with multiple users \cite{He2016Onthe} &  Decoding order, transmission rates, and power &  Instantaneous channel gains of all users and the average channel gain of the eavesdropper  & Target secrecy rate and secrecy outage probability  & A closed-form solution by problem simplification \\
\hline
 Traffic offloading via dual-connectivity in cellular networks \cite{Wu2016Secrecy} & Data rate and transmission power   & Perfect CSI of legitimate channel and  statistics CSI of eavesdropper  &  Traffic demand and  secrecy outage requirement  & Performing a series of equivalent transformations and  proposing an efficient algorithm to compute the optimal offloading solution \\

\hline

\hline

\end{tabular}
\end{table*}

The power consumption of physical-layer secure communications can also be decreased by the designs of secure resource allocation strategies. To be specific, by optimal resource allocation, we can consume as less power as possible to achieve different requirements of secure QoS, as shown in Table \ref{tab_resource_power_min}. The use of AN or jamming signals can deteriorate the wiretap channel, but it also increases the total power consumption. Therefore, the optimal power allocation between the desired information and AN/jamming signals is very important for saving power. In a MISO system in \cite{Romero2012Outage}, the optimal power allocation between transmitted information and AN is developed for minimizing the transmission power while ensuring a given probability of secrecy. In \cite{Akgun2017Exploit} where a multiuser MISO network with friendly jamming is considered, the power allocation strategy is optimized to minimize the total power allocated to the information signals and jamming signals while maintaining secure QoS requirements. A non-orthogonal multiple access system is considered in \cite{He2016Onthe} where a closed-form solution is derived to minimize the transmission power. Additionally, considering the application scenario where an user communicates simultaneously with a macro base station and a small-cell access point, a joint optimization of traffic scheduling and power allocation problem is formulated in \cite{Wu2016Secrecy} with the objective of minimizing the total power consumption while meeting both the user's traffic demand and secrecy requirement.

\subsubsection{Maximization of secure EE}
Secure resource allocation is also effectively used for improving the EE of physical-layer secure communications. To the best of our knowledge, the concept of secrecy capacity per unit cost is defined in \cite{Mustafa2013Per} to study the cost-efficient wide-band secrecy communications, in which the cost of the secrecy capacity may be the number of channel use, the duration of transmission, or the amount of energy consumption. The research status of secure EE maximization by resource allocation can be summarized from the following aspects.

\begin{itemize}
  \item Multiuser multiple-access networks: The secure EE maximization of an OFDMA downlink network is studied in \cite{Ng2012Energy} where the power, secrecy data rate, and subcarrier allocation policies are optimized based on fractional programming and dual decomposition. In a time-division multiple-access network considered in \cite{Butt2015Maximizing}, the secure EE measured by the average energy consumption of the system per transmitted information bit is investigated by using Markov decision process and cross layer design techniques, where information flow and joint optimization of higher and physical layer is permitted. To tackle the problem in \cite{Butt2015Maximizing}, the strategies of packet scheduling and transmitter buffering are designed while the heuristic algorithm of simulated annealing is used to solve the optimization problem due to its advantage to help avoiding local minima.

  \item Multi-antenna networks: The energy-efficient resource allocation is carried out in multi-antenna networks in \cite{Zappone2014Energy} with different CSI scenarios involving perfect CSI, partial CSI, and statical CSI. The work is expanded in \cite{Zappone2016EE} by using the strategy of AN, while the fractional programming and the sequential convex optimization tool are introduced to tackle the nonconvex problem. In \cite{Zappone2017Opt}, based on the optimization framework of \cite{Zappone2014Energy} and \cite{Zappone2016EE}, two EE metrics are optimized, namely the metric of secret bits per Joule and the metric of secret-key EE which is defined as the ratio between the system secret-key capacity and the consumed power. In particular in \cite{Chen2013Energy, Chen2014Energy, Ta2017Adapting}, the optimization problems of energy-efficient secure communications are formulated by using an specific secure EE metric which is therein defined as the ratio of the secrecy outage capacity to the total power consumption.

  \item Relay networks: In \cite{Dong2015Energy, Dong2016Achieving, Dong2016Green}, the energy-efficient power allocation is developed for DF, AF, and untrusted two-way relay networks, respectively. To deal with the nonconvexity of the problems, several optimization approaches are jointly applied, which involve fractional programming, penalty function method, alternating optimization, DC programming. The EE of repetition coding and parallel coding relaying under the partial secrecy regime is investigated by power allocation in \cite{Farhat2016EE} based on the fractional programming and a golden section search algorithm.

  \item Cognitive radio networks: To implement the energy-efficient secure communications in cognitive radio networks in \cite{Gabry2015Energy}, the optimal power allocation and power splitting at the secondary transmitter are optimized under secrecy constraints, while an EE Stackelberg game between the primary and secondary transmitters is formulated for maximizing their utilities. In \cite{Xu2015Energy}, the medium access probability and transmission power of secondary transmitters are jointly optimized to maximize the secure EE of the secondary network. In \cite{Ouyang2016EE}, a secure EE maximization problem is established under the constraints of data rate and transmission power of the cognitive transmission as well as the interference limitation to the primary user, which is solved based on the fractional programming, penalty function method, and DC programming.

  \item The tradeoff  between energy and secrecy: The tradeoff between energy and secrecy also attracts many concerns recently \cite{Comaniciu2013On, Zhang2013Outage, Xu2016SE_EE, Kwon2017D2DEE}. In \cite{Comaniciu2013On} and \cite{Zhang2013Outage}, the tradeoff between energy and secrecy is explored from an information-theoretic perspective, while the metric of partial secrecy is proposed to characterize the secrecy level of a communication system by looking jointly at the application layer metric and physical layer secrecy metric. In \cite{Xu2016SE_EE}, a framework is developed to study the spectrum efficiency (SE) and EE for secure transmission in underlaid random cognitive radio networks, and the joint secure SE and EE optimization problem is formulated therein by using an unified secure SE-EE tradeoff metric. The energy-efficient secure communication in large-scale device-to-device underlaid cellular networks is investigated in \cite{Kwon2017D2DEE}, in which a link adaptation scheme is proposed to strike a balance between secure EE and SE by maximizing the weighted product of secure EE and SE.

\end{itemize}

It is observed that, the most of the secure EE maximization formulations are nonconvex, so that they are very intractable in practice. Therefore, some nonconvex optimization methods are introduced to cope with the challenges, such as the fractional programming, penalty function method, alternating optimization, DC programming, etc. To be specific, the fractional programming can transform the secure EE function (which is a fractional function) into a parameterized polynomial subtractive form which can be tackled by the Dinkelbach algorithm. The penalty function method is able to eliminate the nonconvex constraint of secrecy rate by incorporating the constraint into the objective function. In some cases, the optimization problem is nonconvex or unsolvable for all variables, but it is tractable when we tackle the problem with some of the variables and then tackle it with the rest. Such characteristics are beneficial to implement alternating optimization. Towards the optimization problem in which the objective function can be reformulated as a difference of two convex functions, the DC programming is an effective method which solves the problem iteratively by solving a series of convex subproblems. The explanations of these optimization methods can be found in Table \ref{tab_optimization_methods}.

\renewcommand\arraystretch{1.5}  
\begin{table*}[t]
\centering
\caption{ The explanations of several optimization methods used for secure EE maximization}
\label{tab_optimization_methods}
\begin{tabular}{  m {1.6 cm} m {2.6cm} m {4cm}   m {7.5 cm}   }

\hline

\hline
Optimization methods        &    Problem formulations       &   Problem transformations         &     Algorithm procedures  \\
\hline

\hline
Fractional programming \cite{Dinkelbach, Schaible}
&  \vspace{-5mm}
\begin{equation*}
    \begin{array}{l}
        \max \left\{ f(\bm{x}) = \frac{h(\bm{x})}{g(\bm{x})} \right\} \\
        \text{s.t.}~~ \bm{x} \in \mathbb{D}
    \end{array}
  \end{equation*}  \vspace{-5mm}
& Being related to the parameterized problem $ \max \left\{ h(\bm{x}) -  \varepsilon  g(\bm{x}) : \bm{x} \in \mathbb{D}  \right\} $ with parameter $\varepsilon$.   & \begin{enumerate}
  \item For a given initial value $\bm{x}_0$, calculate $\varepsilon_1 = \frac{h(\bm{x}_0)}{g(\bm{x}_0)} $; let iterative index $i=1$.
  \item For $ \varepsilon_i $, calculate the optimal solution $\bm{x}_i$ by solving the parameterized problem.
  \item Stopping test with $\bm{x}_i$:  If true, then stop; otherwise, go to step 4).
  \item For obtained $\bm{x}_i$, calculate $\varepsilon_{i+1} = \frac{h(\bm{x}_i)}{g(\bm{x}_i)}$, $i:=i+1$, and return to step 2).
\end{enumerate}    \vspace{-5mm}  \\
\hline
Penalty function method \cite{Coleman1982Nonlinear, Bazaraa2013Nonlinear}
&  \vspace{-5mm}
\begin{equation*}
    \begin{array}{l}
        \mathop{\min  f(\bm{x})} \\
        \text{s.t.}~~
        \left\{\begin{array}{l}
        l_k(\bm{x}) \le 0 \\
        \bm{x} \in \mathbb{D},
        \end{array}\right.
    \end{array}
  \end{equation*}  where $k=1,\cdots,m$.
& Defining a penalty function $L(\bm{x})\triangleq \max\{0, l_k(\bm{x})\}_k $ for the nonconvex constraints $l_k(\bm{x}) \le 0$, and transforming the problem  formulation into $\min \{ f(\bm{x})+ \tau L(\bm{x}): \bm{x} \in \mathbb{D} \}$, where $\tau > 0 $ is a penalty factor.
& \begin{enumerate}
  \item Choose a small  penalty factor $\tau_0$ and an increasing factor $\rho$ for updating $\tau$. Let iterative index $i=0$.
  \item For $\tau_i$, calculate the optimal $\bm{x}_{i} $ by solving the resulting penalty problem.
  \item Stopping test with $\bm{x}_{i} $: If true, then stop; otherwise, go to step 4).
  \item Update $\tau$ by $\tau_{i+1} = \rho \tau_i$, $i:=i+1$, and return to step 2).
\end{enumerate} \\
\hline
Alternating optimization  \cite{Gorski2007Biconvex, Niesen2009Adaptive}
&  \vspace{-5mm}
    \begin{equation*}
    \begin{array}{l}
        \mathop{\min  f(\bm{x})} \\
        \text{s.t.}~~
        \bm{x} \in \mathbb{D}
    \end{array}
  \end{equation*} \vspace{-5mm}
& By partitioning the variables $\bm{x}$ into two subsets $\bm{y }$ and $\bm{z }$, the problem can be iteratively solved by tackling the following subproblems
  \vspace{-2mm}
  \begin{equation*}
    \begin{array}{l}
        \!\!\!\mathop{\min\limits_{\bm{y}} \{ f(\bm{y},\bm{z}_i): \bm{y} \!\in\! \mathbb{D}(\bm{y},\bm{z}_i)\} } \\
       \!\!\! \mathop{\min\limits_{\bm{z}} \{ f(\bm{y}_{i+1},\bm{z}): \bm{z} \! \in\! \mathbb{D}(\bm{y}_{i+1},\bm{z}) \} } \\
    \end{array}
  \end{equation*} \vspace{-5mm}
& \begin{enumerate}
  \item Choose a starting point $\bm{x}_0=(\bm{y}_0,\bm{z}_0)$ and let iterative index $i=0$.
  \item For the given $\bm{z}_i$, find the optimal solution $\bm{y}_{i+1}$ of $\mathop{\min\limits_{\bm{y}}  \{f(\bm{y},\bm{z}_i): \bm{y} \in \mathbb{D}(\bm{y},\bm{z}_i) \} }$.
  \item For the given $\bm{y}_{i+1}$, find the optimal solution $\bm{z}_{i+1}$ of $\mathop{\min\limits_{\bm{z}} \{ f(\bm{y}_{i+1},\bm{z}): \bm{z}  \in \mathbb{D}(\bm{y}_{i+1},\bm{z}) \} }.$
  \item Stopping test with $(\bm{y}_{i+1},\bm{z}_{i+1}) $: If true, then stop; otherwise, let $i:=i+1$ and go to step 2).
\end{enumerate}  \\
\hline
DC programming  \cite{Donh2014DC, An2007A}
&  \vspace{-5mm}
    \begin{equation*}
    \begin{array}{l}
     \!\! \!\!  \mathop{\min \{ \! f\!(\!\bm{x}\!)} \!\!=\!\! f_1 \!(\!\bm{x}\!) \!\!-\!\! f_2 \!(\!\bm{x}\!)\! \}\\
      \!\!\!\!  \text{s.t.}~~
        \bm{x} \in \mathbb{D},
    \end{array}
  \end{equation*}
  where $\mathbb{D}$, $f_1 (\bm{x})$, and $f_2 (\bm{x})$  are convex.
& Being solved iteratively by tackling $\min\{f_1 (\bm{x}) - f_2 (\bm{x}_i) - \langle \nabla f_2(\bm{x}_i),  \bm{x}- \bm{x}_i \rangle: \bm{x} \in \mathbb{D} \}$, where $\nabla$ denotes the gradient of a function and $\langle \cdot, \cdot \rangle$ denotes dot product.
& \begin{enumerate}
  \item Choose a starting point $\bm{x}_0$ and let iterative index $i=0$.
  \item For fixed $\bm{x}_i$, find the optimal solution $\bm{x}_{i+1}$ of $\min\{f_1 (\bm{x}) - f_2 (\bm{x}_i) - \langle \nabla f_2(\bm{x}_i),  \bm{x}- \bm{x}_i \rangle : \bm{x} \in \mathbb{D}\}$.
  \item Stopping test with $\bm{x}_{i+1}$: If true, then stop; otherwise, go to step 4).
  \item Let $ i := i + 1$ and go to step 2).
\end{enumerate}

\\
\hline

\hline

\end{tabular}
\end{table*}

\subsection{Secure Beamforming and Precoding}

The deployments of multiple antennas or nodes in wireless networks facilitate the technologies of MIMO or virtual MIMO to be applied extensively, which provide abundant opportunities to perform secure beamforming and precoding \cite{Hong2013Enhancing, Zou2014Improving, Raef2013Coop, Wang2015Enhancing}. It has been demonstrated that, by beamforming and precoding in multi-antenna and multi-node cooperative networks, we can obtain some benefits in terms of secrecy rate, secrecy outage probability, power consumption, and secure EE. Naturally, to gain these benefits, the optimization designs on beamforming and precoding can be solved with the four performance metrics in practice.

\subsubsection{Maximization of achievable secrecy rate}

Following the extensive applications of multi-antenna technologies, secure beamforming and precoding have been paid increasing concerns for secrecy rate improvements \cite{Hong2013Enhancing}. It is verified in \cite{Hero2003Secure} that exploiting space-time diversity at a multi-antenna transmitter can enhance information security and information-hiding capabilities. After that, to improve the secrecy rate of multi-antenna networks, some optimal or suboptimal schemes of secure beamforming/precoding have been proposed for multifarious scenarios based on different methods.

\renewcommand\arraystretch{1.5}
\begin{table*}[t]
\centering
\caption{ The comparisons of the works on the conventional beamforming/precoding schemes}
\label{tab_Beamf_precod}
\begin{tabular}{  p{2.5cm}  p{2cm}  p{3cm}   p{7.5cm}    }

\hline

\hline
 Schemes   &  CSI conditions &  Expression/Constraints   &  Explanations  \\
\hline

\hline
Joint MRT and AN null-space beamforming   \cite{Poor2014Power, Wang2013Secure, Zhang2013On, Zhu2014Secure, Wang2015Secrecy}   &  Legitimate CSI & $\textbf{W}  = \textbf{H}^H / \| \textbf{H} \|$ and $\textbf{HZ}=\textbf{0}$
& Controlling the beam  towards the intended user while Emitting AN  on the null space of legitimate channels. The performance can be improved by power allocation  between AN and information-bearing signal. \\
\hline
ZF beamforming \cite{Wang2013Secure, Wang2012Distributed, Yang2013Cooperative, Dong2010Improving}   &Legitimate and wiretap CSI  &  $\textbf{W} = \textbf{H}^H(\textbf{H}\textbf{H}^H)^{-1} $ or $\textbf{H}_{e} \textbf{W}^H = \textbf{0}$ &Eliminating information leakage to the eavesdropper. This strategy is generally obtained by $\textbf{H}^H(\textbf{H}\textbf{H}^H)^{-1}$ or optimized with the constraint $\textbf{H}_{e} \textbf{W}^H = \textbf{0}$  in system designs. \\
\hline
RCI precoding \cite{Geraci2012Secrecy, Geraci2013Physical, Geraci2013Large, Yang2014Confidential, He2015Base} & Legitimate CSI  & $\textbf{W} = \textbf{H}^H(\textbf{H}\textbf{H}^H + \alpha \textbf{I} )^{-1}  $   &  To balance the intended signal and information leakage by designing a regularization parameter $\alpha$. The secrecy performance of this strategy can be improved by power allocation. \\
\hline
GSVD precoding \cite{Fakoorian2011Dirty, Khisti2010SecureI, Khisti2010SecureII, Fakoorian2011MIMO, Huang2011Cooperative, Fakoorian2012Optimal}  &  Legitimate and wiretap CSI & $\textbf{H}\textbf{W} = \textbf{U}\textbf{A} $ and  $\textbf{H}_e \textbf{W} = \textbf{U}_e \textbf{A}_e $ &  Performing GSVD for matrix $(\textbf{H},\textbf{H}_e)$, and returning the precoding matrix $\textbf{W}$, unitary matrices $\textbf{U}$ and $\textbf{U}_e$, nonnegative diagonal matrices $\textbf{A}$ and  $\textbf{A}_e$. Power allocation can also be optimized for secrecy improvements in this strategy. \\
\hline
\multicolumn{4}{l} {*Notations: $\textbf{H}$, $\textbf{H}_e$, $\textbf{W}$, and $\textbf{Z}$ denote the matrices of legitimate channels, wiretap channels, beamforming/precoding, and AN, respectively.}\\
\hline

\hline
\end{tabular}
\end{table*}

\paragraph{Conventional beamforming/precoding} The conventional beamforming/precoding schemes, such as MRT, signal/AN null space, and GSVD, are applied separately or jointly for secrecy enhancements, due to the inherent simplicity and easy implementation. For achieving a better secrecy performance, power allocation is usually optimized for these schemes. The MRT beamforming controls the beam towards the intended user for strengthening its received signals. Since the MRT beamforming may lead to information leakage on the direction to the eavesdropper, the AN null-space beamforming can then be exploited to disrupt the reception at the eavesdropper by emitting AN on the null space of legitimate channels. Such a joint scheme with MRT and AN null-space beamforming is of particular interest in practice when the eavesdropper's CSI is unavailable. If the transmitter has the full CSI of the eavesdropper, the ZF beamforming can then be performed to overcome the faults of information leakage to the eavesdropper by completely suppressing the beam towards the eavesdropper. To tradeoff the intended received signal and information leakage to eavesdropper or other users, RCI precoding is proposed based on a real regularization parameter which can be designed for secrecy rate improvements. When all nodes in a network are equipped with multiple antennas while the perfect CSI of all nodes is available, the GSVD precoding can be implemented to decompose both the legitimate channels and the wiretap channels into a set of parallel independent subchannels which can be used separately to transmit different messages. The works on the conventional beamforming/precoding schemes are compared in Table \ref{tab_Beamf_precod}. Noteworthily, these conventional beamforming/precoding schemes are suboptimal in many situations, and the optimal designs in this field have therefore attracted great interest.

\paragraph{Optimal beamforming/precoding} To achieve the optimal secrecy performance, the strategy of beamforming/precoding is carefully designed by optimization approaches. The precoding matrix design for maximizing the secrecy capacity $C_s (\textbf{W})$ in a standard three-node (two legitimate users and an eavesdropper) MIMO wiretap network is formulated as \cite{Li2013Transmit}
    \begin{equation}
    \label{SCM_beam}
    \begin{array}{l}
        \mathop{\max \limits_{\textbf{W}} C_s (\textbf{W})}\\
        \text{s.t.}~~
               \text{Tr}(\textbf{W}) \le P^{max}, \textbf{W} \succeq \textbf{0},\\
    \end{array}
    \end{equation}
where $\textbf{W}$ is the precoding matrix with the maximum power constraint $P^{max}$ and the positive semidefinite constraint $ \textbf{W} \succeq \textbf{0}$. Such a nonconvex problem is solved by alternating optimization and dual decomposition, while the resulting algorithm is extended to the scenario with destination jamming. In \cite{Hanif2014On}, the linear precoding strategies for secrecy rate maximization in multiuser multiantenna networks are investigated in the broadcasting and multicasting scenarios, and an iterative algorithm based on second-order cone programming is proposed with  low complexity and provable convergence. Focusing on the secure communications in dual-polarized MIMO systems, a scheme of dual-structured precoding is addressed in \cite{Gong2017Secure} in which a preprocessing matrix based on the polarized array spatial correlation and a linear precoding scheme based on different CSI are concatenated. The secure beamforming for typical three-node (two legitimate users and a relay) MIMO relay networks is explored in \cite{Jeong2012Joint} and \cite{Mo2014Secure}, where the untrusted relay is treated as an eavesdropper. To reduce the difficulties of the joint designs in \cite{Jeong2012Joint} and \cite{Mo2014Secure}, the alternating optimization is used to iteratively deal with the source and the relay beamforming in an alternate fashion. To solve the resulting subproblems from alternating optimization, the SDP is introduced in the both works to transform a fractional quadratically constrained quadratic problem into a SDP problem by the technique of SDR \cite{Luo2010Semidefinite} and the rank-one matrix decomposition theorem \cite{Ai2011New}. Besides, the beamforming for maximizing the secrecy rate in simultaneous wireless information and power transfer is designed in \cite{ Shi2014Secure, Liu2014Secrecy, Khandaker2015Masked}, where the optimal solutions are derived also based on SDR. More specifically, by relaxing the rank-one constraint, the considered optimization problems are therein constructed as SDP problems which can be solved easily by some existing optimization techniques and rank-one reduction \cite{ Shi2014Secure, Liu2014Secrecy, Khandaker2015Masked, Huang2010Rank}.

\subsubsection{Minimization of secrecy outage probability}

In physical layer security, the potential of secure beamforming and precoding for minimizing secrecy outage probability has also been explored in recent years.  Naturally, the existing beamforming/precoding schemes mentioned in the last subsection can also be used to achieve the goal of secrecy outage probability reduction. As in \cite{Xiong2012A}, the AN-assisted beamforming is performed for degrading the eavesdroppers' channels while the optimal power allocation between the confidential information and AN is obtained in closed form to minimize the secrecy rate outage probability. In \cite{Gerbracht2012Secrecy}, the outage probability of secure transmission is minimized by the single-stream beamforming (based on MRT and ZF beamforming) and the use of AN in the null space of the legitimate channels. When only the location information of the eavesdropper is available at the source user in \cite{Yan2016Based, Liu2016Based}, the location-based beamforming is optimally designed to minimize the secrecy outage probability in Rician wiretap channels, while the resulting solution is extended to examine the solution of the optimal beamformer in the presence of a multi-antenna jammer \cite{Liu2016Based}. To transmit information securely in millimeter-wave (mm-Wave) MISO-OFDM systems with partial channel knowledge, a hybrid precoder is implemented in \cite{hybrid2017Ramadan} by an iterative design with the objective  of minimizing the secrecy outage probability.

\subsubsection{Minimization of power consumption}

In the existing literature, secure beamforming and precoding are also used to support the designs of power minimization in different scenarios. The beamforming for minimizing transmission power in relay networks is investigated in \cite{Dong2010Improving, Li2011OnCoop,Liu2014Transmit, Wang2013Joint} with different constraints. The typical mathematical model for minimizing the total power of the source and relays under a target secrecy rate constraint $R_s \ge  R_s^0$ is given as \cite{Dong2010Improving, Li2011OnCoop}
    \begin{equation}
    \label{Relay_beam_Pmin}
    \begin{array}{l}
        \mathop{\min \limits_{P_s, \textbf{w}}\{P_s + \|\textbf{w}\|^2 \}}\\
        \text{s.t.}~~
               R_s (P_s, \textbf{w }) \ge  R_s^0,\\
    \end{array}
    \end{equation}
where $P_s$ and $\textbf{w}$ are the source power and the relay weights, respectively.  In particular in \cite{Wang2013Joint}, the beamformer of the relays is optimized to minimize the power allocated for transmitting confidential information, so that as much power as possible can be used to transmit AN to confuse the eavesdropper. In \cite{JointD2017Kang} where a secure multiuser broadcast system is considered, the optimal precoding matrix at the base station and the jamming covariance matrix at the friendly jammer are jointly designed to minimize the total transmission power under the signal-to-interference-plus-noise ratio (SINR) constraints at the users and eavesdroppers. In \cite{FZhu2016Layer}, the transmission beamforming is performed for minimizing the power consumption of a full-duplex base station considering both self-interference mitigation and physical layer security. Additionally, the physical layer security in satellite communication is considered in \cite{Lei2011Secure} where the beamforming and power allocation under the individual secrecy rate constraints are designed for minimizing the overall transmission power used by all beams. In a new cognitive radio network as described in \cite{Fzhu2016Improv}, a cooperative beamforming scheme is proposed to minimize the transmission power of a secondary transmitter while providing different SINR for an eavesdropper, a primary receiver, and multiple secondary receivers.

The problems of power minimization by beamforming/precoding are also raised in simultaneous wireless information and power transfer systems considering multifarious settings.
\begin{itemize}
  \item Multi-antenna broadcast networks: In such settings, simultaneous wireless information and power transfer is implemented by transmission beamforming which is designed to jointly or separately satisfy the constraints of secrecy rate, secrecy outage probability, energy-harvesting outage probability, and received SINR ratio \cite{OutConst2016Zhu, Zhang2016SecureBeam, zhu2016Beamform, Ng2013ResourceALL, Leng2014Power, Ng2014Robust, Ng2013Multi}. In order to achieve secure transmission, the transmission beamforming is also aided with AN strategy in many works \cite{zhu2016Beamform, Ng2013ResourceALL, Leng2014Power, Ng2014Robust, Ng2013Multi, Zhang2016SecureBeam}.
  \item Distributed antenna systems: In \cite{Ng2015SecureGreen}, the beamforming and AN vectors are jointly optimized to minimize the total transmission power while providing QoS for reliable communication and efficient power transfer in a given time slot, in which the capacity-limited backhaul links is taken into account.
  \item Multi-cell multigroup multicast systems: In \cite{Tervo2017Distributed}, two different optimization targets are considered for a multi-cell multigroup MISO system, i.e., power minimization and SINR balancing. The centralized and distributed beamforming algorithms are proposed for the considered optimization problems, based on the techniques of SDR and alternating optimization.
  \item Cognitive radio networks: Simultaneous wireless information and power transfer are raised in cognitive radio networks in \cite{Zhang2017Coope, Ng2016Multi}. In \cite{Zhang2017Coope}, the total transmission power at the energy transmitter and the secondary transmitter is minimized by a cooperative precoding design while satisfying secrecy rate, energy harvesting, and interference temperature constraints. In \cite{Ng2016Multi}, the total transmission power of the secondary transmitter is minimized while ensuring that the QoS requirement on secure communication is satisfied.
\end{itemize}

It is noted that, in many works, the technique of SDR is extensively adopted in the designs of transmission beamforming \cite{OutConst2016Zhu, zhu2016Beamform, Ng2013ResourceALL, Leng2014Power, Ng2014Robust, Ng2013Multi, Ng2016Multi}, such that an approximation problem can be directly obtained and solved by the method of SDP. In general, the resulting relaxed problem by SDR cannot ensure to get a rank-one solution. It always acts as an upper bound of the performance for the original problem \cite{Luo2010Semidefinite}. In some cases, the solution obtained by SDR is provably optimal, or the rank of the solution can be reduced by some techniques of rank reduction. Noteworthily, solving the SDP problem may result in relatively poor performance if SDP returns a high-rank solution. To overcome the difficulty, a method termed as SPCA \cite{Beck2010Seque, Canelas2017Application} is usually employed to find a suboptimal solution \cite{OutConst2016Zhu, zhu2016Beamform}. The SPCA method approximates the nonconvex constraints by an upper convex estimate, and then results in a problem which can be solved directly. The two methods are briefly compared in Table \ref{tab_SPCA_SDP}.

\renewcommand\arraystretch{1.5}  
\begin{table*}[t]
\centering
\caption{ The comparisons of SDR and SPCA}
\label{tab_SPCA_SDP}
\begin{tabular}{  p  {2.6 cm} p {3 cm}  p  {6cm}  p  {4.3 cm}   }

\hline

\hline
Optimization methods        &    Problem formulations       &   Problem transformations         &     Comments   \\
\hline

\hline
SDR  \cite{Luo2010Semidefinite, Yu1998Semidefinite}
& \begin{equation*}
    \begin{array}{l}
        \min\limits_{\bm{x} \in \mathbb{R}^n} \bm{x}^T \textbf{A}_0\bm{x} \\
        \text{s.t.}~~ \bm{x}^T \textbf{A}_i \bm{x} \ge b_i,
    \end{array}
  \end{equation*}
  where $\textbf{A}_i$ are symmetric square matrices, $i=0,1,2,\cdots$.
& By defining $ \textbf{X} = \bm{x} \bm{x}^T $ which is equivalent to $\textbf{X}$ being a symmetric positive semidefinite matrix with rank one constraint $rank(\textbf{X}) = 1$, we get that $\bm{x}^T \textbf{A}_i \bm{x} = \text{Tr} (\textbf{A}_i  \textbf{X})$. By ignoring $rank(\textbf{X}) = 1$, we obtain a relaxed problem known as an SDP:
  \vspace{-2mm}
  \begin{equation*}
    \begin{array}{l}
        \min\limits_{\textbf{X}} \text{Tr} (\textbf{A}_0  \textbf{X}) \\
        \text{s.t.}~~
        \left\{\begin{array}{l}
            \text{Tr} (\textbf{A}_i \textbf{X}) \ge b_i,  i = 1,2,\cdots \\
            \textbf{X} \succeq 0.
        \end{array}\right.
    \end{array}
  \end{equation*} \vspace{-3mm}
& The core idea of the method is that we drop the rank-one constraint to obtain a SDP problem. The SDP problem can be handled very conveniently by readily available software packages. However, the resulting SDP problem may lead to relatively poor performance if SDP returns a high-rank solution. \\
\hline
SPCA \cite{Beck2010Seque, Canelas2017Application}
&  \vspace{-3mm}
   \begin{equation*}
    \begin{array}{l}
        \min\limits_{\bm{x} \in \mathbb{R}^n} f(\bm{x}) \\
        \text{s.t.}~~\!\!
            l_i (\bm{x}) \!\le\! 0, i \!=\! 1,2,\cdots
    \end{array}
  \end{equation*}
   where $f(\bm{x})$ is convex, and $l_i (\bm{x})$ is nonconvex.
& By defining a function $L_i (\bm{x}, \varphi_i )$ which is a convex upper approximation of the nonconvex function $l_i (\bm{x})$, i.e., $l_i (\bm{x}) \le L_i (\bm{x}, \varphi_i )$, the original problem can be approximated by the following convex problem:
   \vspace{-2mm}
   \begin{equation*}
    \begin{array}{l}
        \min\limits_{\bm{x} \in \mathbb{R}^n} f(\bm{x}) \\
        \text{s.t.}~~
            L_i (\bm{x}, \varphi_i) \le 0, i =1,2,\cdots,
    \end{array}
  \end{equation*}
   where $\varphi_i$ is a slack variable which is updated at each iteration.
& The basic idea of the method is that, at each iteration, we replace each of the nonconvex constraints by its upper convex approximation function with an  appropriate $\varphi_i$. Thus, the method is required to iteratively solve a convex problem based on convex optimization. The difficulty of the method is to carefully choose the upper convex estimates and slack variables. \\
\hline

\hline

\end{tabular}
\end{table*}

\subsubsection{Maximization of secure EE}

The energy-efficient beamforming and precoding in physical layer security have also been given many attentions. In \cite{Zhang2014Energy}, the energy-efficient precoder design in a conventional three-node (including a transmitter, a legitimate receiver, and an eavesdropper) MIMO wiretap channel is proposed based on the fractional programming and Taylor series expansion. In \cite{Gursoy2012Secure}, by providing a second-order approximation to the MIMO secrecy capacity with its first and second derivatives, the metric of minimum bit energy is examined for secure and reliable communications in the low-SNR regime while characterizing  the tradeoff between EE and secrecy. A beamformer design is performed in \cite{Nghia2017MIMO} for secure and energy-efficient wireless communication over MIMO channels with multiple user pairs and an eavesdropper, where a path-following computational procedure is proposed to cope with the intractable nonconvex problem and to yield at least a locally optimal solution. In \cite{Mei2016ROBUST}, the robust energy-efficient transmission design for MISOME wiretap channels is investigated by the fractional programming and tight convex relaxation, so that the primal fractional optimization problem is solved by solving a sequence of SDP problems. The energy-efficient beamforming for secure cognitive communication is raised in \cite{Ouyang2016SEE}, in which the primal problem is tackled by the combined use of the fractional programming and DC programming. In addition, in a MIMOME network with simultaneous wireless information and power transfer \cite{Mei2017Artif}, the transmission covariance matrices and power splitting ratio for decoding information and harvesting energy are designed jointly to maximize the secure EE, where the fractional programming and alternating optimization are also employed for handling the nonconvexity of the optimization problem.

\subsection{Antenna/Node Selection and Cooperation}

Antenna/node selection and cooperation in multi-antenna and multi-node wireless networks have been well exploited for achieving different performance requirements of physical layer security. Being similar to the former subsections, the state of the art of optimization designs in this research topic can also be reviewed from the four categories of optimization problems.

\subsubsection{Maximization of achievable secrecy rate}
Great efforts have been made for the optimization designs of antenna/node selection and cooperation to increase the achievable secrecy rate. Multi-antenna diversity can provide the gain of secrecy rate by designing proper strategy of antenna selection, as investigated in \cite{Hanif2013Efficient, Yan2014Transmit, Alves2016Enhanced}. In multiuser scenarios, user selecting/scheduling can bring the improvement of secrecy rate by using multiuser diversity, such as the optimal and suboptimal scheduling in a multiuser MISO system \cite{Krikidis2013Secrecy}, the maximum instantaneous SNR scheduling and approximate proportional fair scheduling in a multiuser MISO system with a multi-antenna eavesdropper \cite{Pei2014On}, and the round-robin user scheduling as well as the optimal and suboptimal user scheduling in a cognitive radio network \cite{Zou2013Physical, Zou2014Secrecy}.

\begin{figure*}[t]
\centering
\subfigure[Intermediate nodes used as relays for beamforming]{
    \label{a_relaying}
    \includegraphics[width=0.4\textwidth]{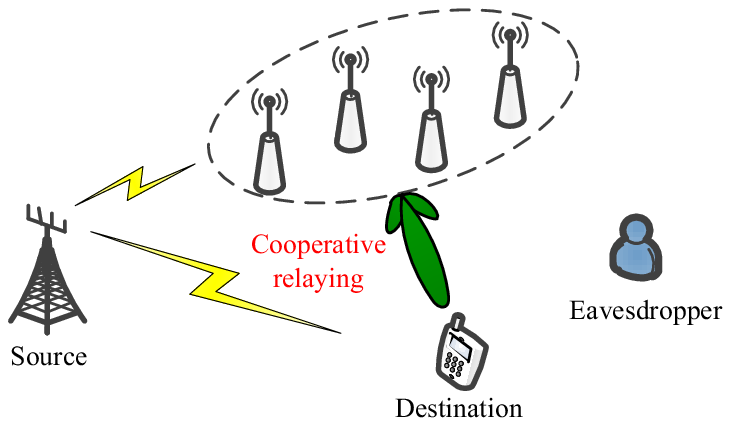}}
\hspace{1cm}
\subfigure[Intermediate nodes used as jammers for beamforming]{
    \label{b_jamming}
    \includegraphics[width=0.4\textwidth]{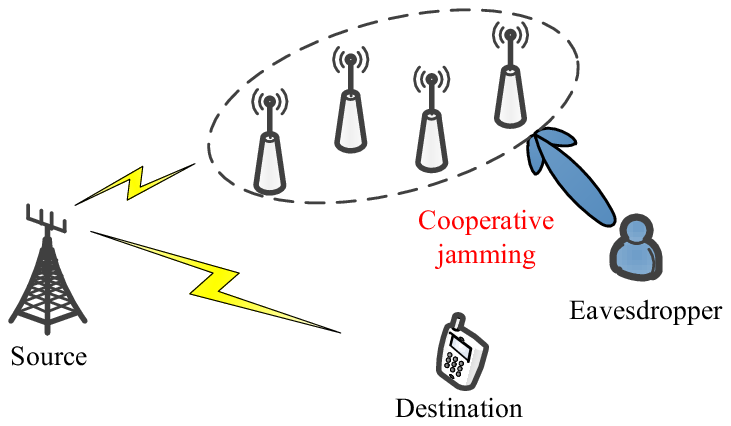}}
\subfigure[Intermediate nodes grouped as relays and jammers for beamforming]{
    \label{c_relaying_jamming}
    \includegraphics[width=0.4\textwidth]{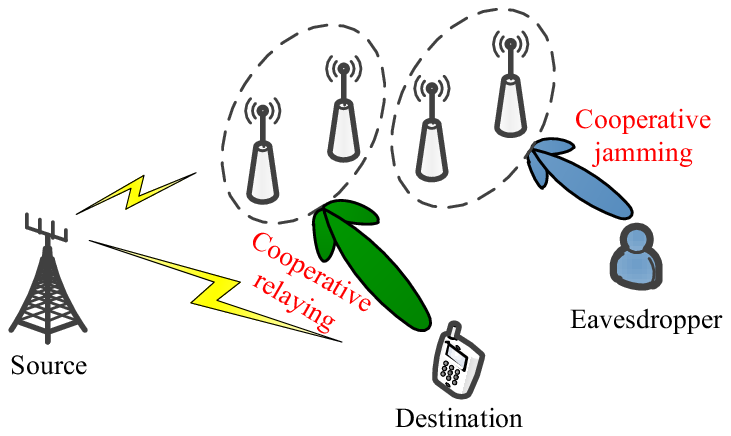}}
\hspace{1cm}
\subfigure[Intermediate nodes used for joint relay beamforming and AN precoding]{
    \label{d_relaying_AN}
    \includegraphics[width=0.4\textwidth]{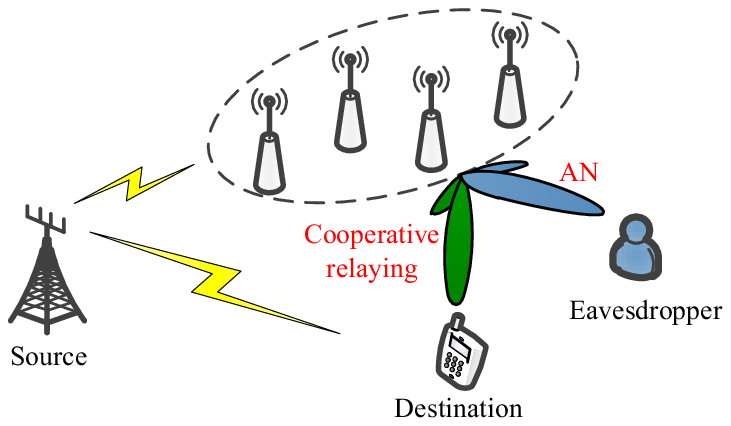}}
\caption{Different strategies of intermediate node assisted transmission in  cooperative networks.}
\label{Fig_Intermediate}
\end{figure*}

In cooperative networks, the broadcast feature of wireless transmission results in two aspects, namely node cooperation and data secrecy \cite{Trappe2010wireless}. Node cooperation means that users can help improve each other's rate by intelligently using their received signals. Data secrecy implies that the information leakage to the undesired users may cause some severe problems of information security. Accordingly, node cooperation and data secrecy have been studied jointly in multi-node cooperative scenarios in recent years. The cooperative nodes act two roles in physical layer security, including cooperative relaying and cooperative jamming \cite{Lai2008Relay, Hong2013Enhancing, Mukh2014Principles, Wang2015Enhancing, Yener2015Wireless, Rod2015Physical}. Cooperative relaying is to enhance the legitimate channels while cooperative jamming is to degrade the wiretap channels. In practice, the cooperative nodes may be trusted or untrusted. For the trusted nodes, they can be used for relaying and jamming separately or jointly \cite{Rod2015Physical}. As to untrusted nodes, seeking for cooperative relaying or jamming with them may be better than treating them as pure eavesdroppers \cite{He2010Cooperation, He2008The}. According to the roles of the cooperative nodes, there are generally four kinds of node-assisted transmission designs, which involve cooperative relaying, cooperative jamming, hybird cooperative relaying and jamming, and cooperative relaying with AN \cite{Wang2015Enhancing, Li2015RobustCoop}, as illustrated in Fig. \ref{Fig_Intermediate}.
\begin{itemize}
  \item Cooperative relaying: When the channels from the source to the destination are too poor or even nonexistent, signal retransmission by intermediate nodes is an effective way for confidential data transmissions, as shown in Fig. \ref{a_relaying}. Seeking for cooperative relaying with the intermediate nodes, the confidential data can be delivered securely and reliably, while some signal processing technologies can be applied into system designs to achieve both the performance requirements and resource saving. The typical cooperative relaying supported by beamforming to improve secrecy rate is investigated in \cite{Dong2010Improving, Li2011OnCoop, Wang2012Distributed, Jeong2012Joint, Wang2013Secure, Yang2013Cooperative, Vishw2013Decode, Mo2014Secure}, where the relays are trusted \cite{Dong2010Improving, Li2011OnCoop, Wang2012Distributed, Wang2013Secure, Yang2013Cooperative, Vishw2013Decode, Zheng2015Outage} or untrusted \cite{Jeong2012Joint, Mo2014Secure}. The optimal power control for multi-hop relaying is raised in \cite{Lee2016Optimal}. The optimal relay selection and relay placement for cooperative relaying are concerned in \cite{Zou2013Optimal} and \cite{Mo2012Relay}, respectively. In \cite{Bao2013Relay}, three opportunistic relay selection schemes are studied for maximizing the Shannon capacity to the destination as well as for minimizing that to the eavesdroppers. According to \cite{Bao2013Relay} and \cite{Krikidis2009Relay}, the relay selection schemes can be sorted into four categories, as listed in Table \ref{tab_relay_selection}.

\begin{table*}[t]
\centering
\caption{Relay selection schemes}
\label{tab_relay_selection}
\begin{tabular}{  p{2.8cm}  p{3.2cm}  p{3.8cm}   p{6cm}   }

\hline

\hline
 Schemes   &   Mathematical formulations  &  Achievable secrecy rates     & Explanations \\
\hline

\hline
 Conventional selection &  $ k^* = \arg \max \limits_{k\in \Omega} \left\{\gamma_{k,D}\right\}$  &  $C_s = \max\limits_{k \in \Omega} \{ C_{k,D}\} - C_{k^*, E}$  & The relay which has the highest instantaneous SNR of the relay-destination links  will be selected for relaying.  \\
\hline
Minimum selection  &   $ k^* = \arg \min \limits_{k\in \Omega} \left\{\gamma_{k,E} \right\}$  &  $C_s = C_{k^*,D} - \min \limits_{k \in \Omega} \{ C_{k,E}\}$ &   The relay that has the lowest instantaneous SNR of the relay-eavesdropper links will be selected for relaying. \\
\hline
Optimal selection  &  $k^* = \arg \min \limits_{k\in \Omega} \left\{ \frac {1+\gamma_{k,D}}  {1+ \gamma_{k,E}}  \right\} $   &   $ C_s = C_{k^*,D} -  C_{k^*, E}$  & The optimal selection incorporates the quality of both links into the selection decision metric. The relay that has the highest achievable secrecy rate will be  selected for relaying.\\
\hline
Suboptimal selection & $k^* \!=\! \arg \min \limits_{k\in \Omega} \left\{ \frac {\gamma_{k,D}}  { \mathbb{E} \left\{\gamma_{k,E}\right\}}  \right\} $ & $ C_s = C_{k^*,D} - \mathbb{E} \left\{ C_{k^*, E}\right\}$  &  The suboptimal selection scheme selects the appropriate relay based on the statistical knowledge of the relay-eavesdropper links. The scheme can avoid the instantaneous estimate of the wiretap channels. \\
\hline
\begin{minipage}{17cm} \vspace{1mm} *Notations: $C_s$ is the secrecy rate. $k^*$ is the selected relay from the relay set $\Omega$. $\gamma_{k,D}$ and $\gamma_{k,E}$ are the instantaneous SNR of the links from relay $k$ to the destination and the eavesdropper, respectively. $C_{k,D}$ and $C_{k,E}$ are the channel capacity of the links from relay $k$ to the destination and the eavesdropper, respectively. $\mathbb{E}\{\cdot\} $ denotes expectation. \vspace{1mm} \end{minipage}\\
\hline

\hline

\end{tabular}
\end{table*}

    \item Cooperative jamming: When there is the direct channel from the source to the destination, the relays can be used as jammers to emit artificial interference, such that the channels to the eavesdropper are degraded and the confidential information is protected against eavesdropping, as depicted in Fig. \ref{b_jamming}. A simple but suboptimal jamming strategy is null-space cooperative jamming which emits artificial interference in the null space of the channels from the relays to the destination. Such a null-space jamming strategy degrades only the wiretap channels while with no influence to the legitimate channels. Trying to obtain the optimal cooperative jamming designs, the solutions of jamming signal weights are elaborated in \cite{Dong2010Improving, Yang2014Joint }, and \cite{Li2011OnCoop}. In \cite{Huang2011Cooperative}, two types of cooperative jamming schemes referred to as full cooperative jamming and partial cooperative jamming are proposed depending on that whether both the transmitter and the temporary helpers (which are acted by the source and the destination) transmit jamming signals at the same time. The secure transmissions with and without cooperative jamming are compared in \cite{Huang2012Robust} based on the worst-case optimization. Using the intermediate nodes to relay or to jam, which is a better choice? Such a problem involving cooperative mode decision is discussed in \cite{Deng2015Secrecy}, in which the performance comparison between the relay transmission scheme and the direct transmission scheme with jamming is examined in the distance normalized SNR regime. Additionally, in \cite{Deng2015Secrecy}, it is also concluded that, in the high distance normalized SNR regime, the direct transmission scheme provides higher secrecy rate compared with the relay transmission scheme, while in the low distance normalized SNR regime, the relay transmission scheme outperforms the direct transmission scheme.

    \item Hybrid cooperative relaying and jamming: A more widely-used strategy is hybrid cooperative relaying and jamming which is based on the combined application of the two methods. In such a hybrid strategy, the intermediate nodes are grouped as relays and jammers. The relays retransmit the received confidential information to improve the signal quality at the destination while the jammers continuously emit artificial interference to confuse the eavesdroppers, as illustrated in Fig \ref{c_relaying_jamming}. This may take the confidential information under protection in all stages of cooperative transmission. As in \cite{Krikidis2009Relay, Chen2011Joint, Kim2012Combined, Wang2013Hybrid, Wang2015Hybrid, Zhang2015Partner}, the hybrid cooperative relaying and jamming are investigated in different settings where joint relay and jammer selection are also taken into account. A multiuser relaying scheme with the intended user jamming is proposed in \cite{Kim2015Secure} where the optimal user selection is also considered in the sense of maximizing the secrecy rate. In \cite{Zhang2012Physical}, the untrusted two-way relaying with friendly jammers is investigated based on Stackelberg game. In \cite{Liu2013Destination}, a hybrid relaying and jamming scheme with the optimal relay selection and power allocation is developed for maximizing the secrecy rate, in which the destination and the source are used as jammers to jam the eavesdropper in the first and second phase of cooperative transmissions, respectively.

    \item Cooperative relaying with AN: In this strategy, as shown in Fig. \ref{d_relaying_AN}, the relays are used to forward confidential information and transmit AN simultaneously. To be specific, the confidential information retransmitted at each relay is superimposed with an AN. This strategy subsumes all the above three designs and makes better use of available degrees of freedom at relays \cite{Li2015RobustCoop}. As a result, the combined designs of cooperative relaying and AN precoding have been widely considered in physical layer security, such as in \cite{Wang2014RobustJ, Li2015RobustCoop, Yang2013Optimal, Lin2014Joint}. Noteworthily, it is generally challenging to jointly optimize AN precoding and relay beamforming \cite{Li2015RobustCoop}. Therefore, some suboptimal but effective schemes have been proposed. The joint AN-aided beamforming and power allocation are studied in \cite{Huang2011Cooperative}, where a closed-form jamming beamformer and a GSVD-based secure relaying scheme with each corresponding optimal power allocation are developed for the cases of single and multiple stream data transmissions, respectively. In \cite{Ding2012A}, for a single-relay MIMO wiretap channel, an interference alignment approach is addressed to obtain a simplified and suboptimal design of AN-assisted cooperative relaying.

\end{itemize}

\subsubsection{Minimization of secrecy outage probability}

The spatial diversity provided by multiple antennas and nodes can be effectively utilized for reducing the secrecy outage probability in multi-antenna and multi-node networks. The antenna selection for enhancing the secrecy outage performance has been investigated for MIMOME networks \cite{Hu2015Secrecy, Huang2015SecureTrans}, MIMO relay networks \cite{Zhao2017SecrecyPerf}, non-orthogonal multiple access systems \cite{Lei2017OnNOMA}, and cognitive radio networks\cite{Chen2015Dual}. The antenna selection combined with AN is proposed for decreasing the secrecy outage probability in secure two-way relaying communications in \cite{Ding2014Asymptotic}. In multi-node scenarios,  the relay and jammer selection in DF relay networks are studied separately or jointly for minimizing the secrecy outage probability in \cite{Wang2015Generalized, Hui2015Secure}. The best relay and user pair selection for minimizing the secrecy outage probability of a multiuser AF relay network are addressed in \cite{Fan2014Secure}, in which the analytical expressions of the secrecy outage probability are also derived for the proposed three selection criteria. The selections of the transmission protocols are also investigated in literature \cite{Lin2016Rob, Vien2017Oppo}. In \cite{Lin2016Rob}, the secure transmission protocol which switches between DF beamforming and cooperative jamming is proposed for the purpose of maximizing secrecy rate and minimizing secrecy outage probability in different communication scenarios. A secure opportunistic transmission protocol that aims at finding an optimal scheme between direct transmission and relaying transmission, is developed in \cite{Vien2017Oppo} to achieve the lowest secrecy outage probability of cooperative wireless networks. Some works also consider designing the strategies of node selection and cooperation to reduce the secrecy outage probability of cooperative energy harvesting networks \cite{Hoang2017Phys, Moon2016Secrecy, Secrecy2017Moonlee}. In \cite{Hoang2017Phys}, the relay and jammer selection are considered for the cooperative energy harvesting networks with a friendly jammer. In \cite{Moon2016Secrecy}, the secrecy outage probability of a wireless powered communication network with an energy harvesting jammer is analyzed and minimized by optimizing the time allocation between the two phases of information transfer and energy transfer. The work presented in \cite{Moon2016Secrecy} is extended to a more general multiuser situation with an additional consideration of jamming power allocation in \cite{Secrecy2017Moonlee}.

\subsubsection{Minimization of power consumption}

Although multiple node cooperation can support the improvements of information security, multiple nodes used for information transmission may bring additional power consumption. In particular, some cooperative nodes may consume high power but bring inconsiderably improvement of secrecy. Accordingly, node selection and cooperation for saving power while ensuring secure QoS requirements have been also studied in physical layer security. As investigated in \cite{Nomikos2015Relay}, the so-called power-efficient secure communication is discussed with the objective of power minimization by optimal relay selection. In \cite{Mabrouk2017ANAided}, a secure adaptive relay cooperation approach is developed to ensure wireless information security in an untrusted relay network with relay energy harvesting, while a greedy battery-aware relay selection scheme is proposed to minimize the power consumption in such a network.

\subsubsection{Maximization of secure EE}

It has been verified that antenna/node selection and cooperation also can bring the gain of secure EE. In \cite{Farhat2017SEEMRC}, the secure EE of a cooperative MIMO relay network is investigated, in which transmit antenna selection and MRC are deployed at the transmitter and the receivers, respectively. Considering three possible cooperation scenarios in \cite{Wang2014Adaptive}, namely the jammer only, relay only, and the relay-jammer pair, the adaptive cooperation schemes are addressed for energy-efficient physical layer security. In \cite{Zheng2017PhyLa}, hybrid full-/half-duplex receiver deployment strategies are proposed for wireless ad hoc networks to optimize the network-wide secrecy throughput and network-wide secure EE, respectively. The potential advantages of massive MIMO technologies are also explored for improving secure EE \cite{Wang2016Secrecyand, Chen2017Green}. In \cite{Wang2016Secrecyand}, the potential benefits of massive MIMO aided heterogeneous cloud radio access networks are explored in terms of the secrecy and EE. In \cite{Chen2017Green}, the advantages of massive MIMO relaying are utilized to improve the secure EE which is specially defined as the ratio of the secrecy outage capacity to the total power consumption. Moreover, the energy-efficient secure communication over a large-scale wireless network is studied by the combined application of game theory and stochastic geometry in \cite{Kwon2017Game}. An alternating optimization scheme is proposed therein for maximizing the secure EE of the legitimate transmitters by controlling the node activation probability, confidential message rate, redundancy rate, and the number of active antennas. In addition, an energy-efficient node activation game between the transmitters and the eavesdroppers is also studied therein, where the transmitters and the eavesdroppers control their node activation probabilities to maximize the secrecy EE and the eavesdropping EE, respectively.


\section{The impacts of CSI on physical-layer security designs}\label{section_CSI}

It has been discussed that the priori knowledge of the legitimate and wiretap channels' CSI is very important for the choices of secrecy metrics and the designs of secrecy strategies \cite{Wang2015Enhancing}. To achieve the optimal performance of secure transmission, the perfect CSI of both the legitimate and wiretap channels is indispensable for system designs. For getting the CSI of the legitimate channels, some conventional methods (such as training/estimation and feedback), being similar to that in the traditional communications without secrecy constraints, can be used in physical layer security designs. However, due to the existences of estimation error and feedback delay in some cases, it may be difficult in practice to get the perfect CSI of legitimate channels. Regarding the CSI of wiretap channels, it can be obtained perfectly when the eavesdroppers are also the legitimate users of the network but have different service from that of the intended users. However, when the eavesdroppers are passive, vicious or even hostile, it may be impossible to get the perfect CSI of such eavesdroppers. According to the above discussions, the following assumptions of CSI have been considered in physical layer security, i.e., the perfect CSI of all channels, the imperfect CSI of wiretap channels, and the unknown CSI of wiretap channels.

\subsection{The Perfect CSI of All Channels}

In the literature on physical layer security, the perfect CSI of all channels has been commonly assumed for designing the optimal transmission scheme which can match the instantaneous changes of channel states, such as in \cite{Dong2010Improving, Li2011OnCoop, Tao2011Subcarrier, Jeong2011Optimal, Lee2016Optimal, Jeong2012Joint, Lv2015Secrecy, Wang2012Distributed, Zou2013Optimal, Yang2013Cooperative}. In fact, the perfect CSI including that of eavesdroppers, can be obtained at all communication nodes in some situations. For instance, the eavesdropper is active in the network and its transmissions can be monitored. This case arises particularly in the practical applications combining multicast and unicast transmissions, in which the user plays double roles as legitimate receiver for some signals and eavesdropper for others \cite{Dong2010Improving}. Alternatively, the eavesdropper is also a legitimate user of the network whereas its service differs from that of the intended user \cite{Yang2013Cooperative}. In other words, instead of eavesdroppers, there can be friendly nodes in the network that are not supposed to hear certain messages. This case arises often in military communications, where lower level network users can only access to less information \cite{Li2011OnCoop}. Because the confidential information of the source user is expected to be received only by the intended user, the other users (they are even legitimate and friendly) in the network should be treated as eavesdroppers for secure transmission designs. However, such legitimate and friendly users can feed back the perfect CSI to transmitters. Accordingly, the optimal secure transmission designs can be performed with the perfect CSI of all channels.

\subsection{The Imperfect CSI of Wiretap Channels}

In many situations, the perfect CSI of the main channel can be easily obtained by channel estimation and CSI feedback, whereas getting the perfect CSI of the wiretap channels is very difficult or even impossible. In such cases, the imperfect CSI of eavesdroppers may be obtained in practice, based on the past channel observations or a priori knowledge of the particular propagation environment \cite{Jiangyuan2011Coop, Goldsmithl2003Capa}. The uncertainties of the imperfect eavesdropper's CSI can be generally characterized by three ways. The first way is that the channel of eavesdropper follow some probability distributions \cite{Zheng2015Outage, Wang2015Hybrid}, such as the Gaussian distribution, Rayleigh distribution, Rician distribution, and so on. In this way, only the statistical information of the eavesdroppers' channels, i.e., the mean and covariance of the probability distribution, is available for the system designs, such as the assumptions in \cite{Poor2014Power, Wang2013Secure, Dong2015Energy, Ng2012Energy, Park2013On}. The second way to characterize the uncertainties of eavesdroppers' channels is termed as the deterministic uncertainty model in some literature \cite{Wang2013Secure, Huang2012Robust, Wang2014RobustJ, Li2015RobustCoop, Ng2014Robust, Ma2013outage, Lei2011Secure }. In the deterministic uncertainty model which belongs to compound channel in information theory, the unknown wiretap channels are assumed to fall in a sphere or a set. To be specific, the uncertainty region of eavesdropper's channels is modeled as a sphere $\mathbb{H}_e$ with center $\bar{\mathbf{h}}_e$ and radius $\sqrt{\epsilon}$, that is \cite{Wang2013Secure, Huang2012Robust, Wang2014RobustJ, Li2015RobustCoop, Ng2014Robust, Ma2013outage, Lei2011Secure }
    \begin{equation}
    \label{uncertainty}
    \begin{aligned}
         \mathbb{H}_e &= \left\{\mathbf{h}_e | \| \mathbf{h}_e - \bar{\mathbf{h}}_e \|^2 \le \epsilon  \right\}  \\
                                             &= \left\{ \bar{\mathbf{h}}_e + \mathbf{v}_e | \| \mathbf{v}_e\|^2 \le \epsilon  \right\}.
    \end{aligned}
    \end{equation}
In \eqref{uncertainty}, $\mathbf{h}_e$, $\bar{\mathbf{h}}_e$, $\mathbf{v}_e$, and $\epsilon > 0$ denote the real channel vector of eavesdropper, the estimated channel vector of eavesdropper, the estimation error vector, and the channel mismatch, respectively. By this model, we have that ${\mathbf{h}}_e \in \mathbb{H}_e$. The third way to model the imperfect eavesdropper's channels is based on the imperfect channel estimate $\bar{\mathbf{h}}_e$, the estimation error vector $\mathbf{v}_e$, and a scalar $\kappa \in [0, 1]$ for indicating the degree of channel knowledge. This model can be expressed as \cite{Gerbracht2012Secrecy, Chen2015LargeMIMO}
    \begin{equation}
    \label{uncert_model}
    \begin{aligned}
          \mathbf{h}_e = \sqrt{\kappa} \bar{\mathbf{h}}_e + \sqrt{1-\kappa} \mathbf{v}_e.
    \end{aligned}
    \end{equation}
In \eqref{uncert_model}, if $\kappa =1 $, it means that the eavesdropper's CSI is perfect, while if $\kappa =0 $, it implies that we fail to get any CSI of the eavesdroppers.

In some worse cases, the perfect CSI of both legitimate and wiretap channels is unavailable due to limited feedback or other reasons, such as discussed in \cite{Park2013On, Zheng2015Optimal, Zhao2015Robust}. Then, the uncertainties of legitimate channels can also be characterized by the three methods mentioned above. It is worth noting that, towards the uncertainties of real channels, the robust secure designs are commonly performed to ensure achieving the security, reliability, and robustness of information transmission \cite{Zhao2015Robust, Ma2013outage, Wang2014RobustJ, Li2015RobustCoop, Huang2012Robust, Ng2014Robust}.




\subsection{The Unknown CSI of Wiretap Channels}

The assumption on the perfect CSI of all channels is commonly used for calculating the instantaneous secrecy capacity and secrecy rate which are needed for instantaneous optimization designs. Using the perfect CSI, the security and reliability of information transmission can be guaranteed by secure coding and rate adaptation. However, a more practical assumption is that the CSI of wiretap channels is completely absent due to the concealment and hostility of eavesdroppers \cite{Wang2013Joint, Zhang2016Secrecy, Glauber2015Secrecy, Huang2015SecureTrans}. Moreover, whether there exists any eavesdropper cannot be known in some situations. Because the eavesdroppers' CSI is unknown at the transmitters, the expression of the instantaneous secrecy rate is unavailable. Therefore, the instantaneous optimization cannot be performed. Then, a probabilistic view of security or a QoS-based optimization can be considered for secure transmission designs. Such as in \cite{Huang2015SecureTrans}, a strategy of transmission antenna selection to enhance the secrecy performance of MIMO wiretap channels without eavesdroppers' CSI is proposed based on three important metrics, i.e., the probability of non-zero secrecy capacity, the secrecy outage probability, and the $\epsilon$-outage secrecy capacity. In \cite{Zhang2016Secrecy}, secrecy sum rate maximization considering each user's QoS constraint and unknown eavesdropper's CSI is investigated for a non-orthogonal multiple access system. In \cite{Wang2013Joint}, a QoS-based secure strategy is addressed to enhance the security of a cooperative relay network without eavesdropper's CSI. It is worth pointing out that, exploiting AN or jamming signal to enhance secrecy has been demonstrated to be effective when the eavesdropper's CSI is unknown or imperfect \cite{Hong2013Enhancing}.


\section{Discussions on Future Directions and Challenges}\label{section_Direction}

It has been shown in previous sections that the physical layer security has attracted increasing concerns. Some great progress has been made in the fields of information-theoretical security and optimal secure designs at physical layer. However, it has been observed that many studies in the existing works are performed with some special assumptions on CSI, eavesdropper model, and application scenarios. These assumptions may be unpractical or even contrary to real conditions. Therefore, there are still many significant problems needed to be investigated to promote the practical applicability of physical layer security. In the following, some possible future directions and open challenges are simply discussed. Since the future work in physical layer security is very extensive, only a few directions are discussed.

\subsection{The Influences of Wireless Channels}
The influences of wireless channels on secrecy must be further studied. In literature, it is often assumed that the channels to legitimate user and eavesdropper are uncorrelated. The uncorrelated property is believed to be the foundation to assume that the eavesdroppers cannot estimate the channels of legitimate transceivers. However, this assumption has its limitations considering some practical scenarios. For instance, when the transceivers as well as the eavesdroppers lie in a insufficiently rich scattering environment as discussed in \cite{WadeTrappe2015Cha}, the assumption of uncorrelated channels is then impractical. In addition, much existing literature simply assume that the channels are quasi-static or even completely static. However, if the channels are somewhat dynamic, the resulting conclusions in those works may be in conflict with the real settings. Furthermore, the relative spatial locations between the transceivers and eavesdroppers, as well as the node mobility model, may have important impacts on wireless channels, which also need to be considered in secure transmission designs. Besides, it is already known that the secure strategy designs heavily depend on the CSI of legitimate users and eavesdroppers, whereas the perfect CSI is difficult to get in many situations due to the limited estimation and feedback or other reasons.

The challenges stemmed from the aspect of wireless channels are because of the difficulties of accurate channel estimation for wiretap channels, and the considerations of channel correlations, time varying, and node mobility. First, how to get the perfect CSI to achieve the optimal security performance is difficult to deal with, especially when the eavesdropper is inactive. Furthermore, accurate channel estimation may cause unacceptably high
overhead in pilot frequency and power consumption. This is a particularly severe problem in massive MIMO networks as the overhead may grow rapidly with the antenna number. Even worse, the process of channel estimation may be attacked by pilot contamination attack which not only dramatically reduces the achievable secrecy capacity but is also difficult to detect \cite{Kapetanovic2015Physical}. Second, high channel correlations have been observed in \cite{Kyr2006CorreMIMO} even when the spatial separation is much larger than half-wavelength \cite{linksignature2014}. This indicates that the spatial correlations of wireless channels may vary in different environments and the half-wavelength decorrelation assumption may not always hold \cite{linksignature2014}. Therefore, the secure transmission designs considering the channel correlations is also a challenging problem in future. Third, the time-varying characteristics of channels and the mobility of terminals are also severe issues in physical-layer secure communications since the channel qualities may vary dramatically over time and space. Therefore, how to simultaneously guarantee the security, reliability, and robustness of a secure transmission scheme with the problems mentioned above will be challenging in future work.


\subsection{The Impacts of Adversary Model}
The impacts of attack modes and adversary models are also important issues for secure transmission that has not yet been deeply explored. Much existing literature assumes that the adversaries merely passively listen to the secure communications. In other words, there are no collaboration and information exchange among the adversaries. Nevertheless, the adversaries may actively collaborate and exchange their outputs in practice to interpret the confidential messages \cite{Secure2011Liang}. Moreover, a slightly more sophisticated adversary may be able to predict the channels for improving the eavesdropping qualities. Some intelligent adversaries may attempt to manipulate the propagation environment for strengthening their advantages and undermining information security \cite{Xioa2017combating, ABEDI2017robust}. When these observations discussed above are taken into account, the transmission strategy designs for physical layer security will be facing great challenges.

The challenges in this direction can be discussed from the following aspects. On the one hand, the optimization and design in physical layer security will become more complicated when hybrid attacks are imposed on wireless information transmission, such as eavesdropping attack, jamming attack, denial-of-service attack, spoofing attack, message falsification/injection attack, etc. It will be of particular importance to develop new techniques to jointly defend against hybrid wireless attacks \cite{Zou2016security}. On the other hand, great difficulties result from the intelligent adversaries that not only can efficiently collaborate with each other and actively manipulate propagation environment for attacks, but also can autonomously learn the knowledge of the associated wireless network to find its weakness and then to implement adaptive attacks. Therefore, it is challenging to develop well-performing secure mechanisms to
defend against the intelligent adversaries.

\subsection{The Influences of Hardware Impairments}

Hardware impairments are nonnegligible factors which should also be taken into account in physical layer security. So far, a great deal of works on the designs of security strategies assume that the transceiver hardware is perfect. However, hardware impairments truly exist in practice, due to nonlinear power amplifiers, in-phase and quadrature (I/Q) imbalance, frequency and phase offsets, quantization noise, and synchronization errors \cite{Boshkovska2017Power}. For instance, I/Q imbalance can attenuate the amplitude and rotate the phase of the desired constellation, while it can create an additional signal from the mirror subcarrier which leads to a symbol error rate. In the presence of nonlinearities of power amplifiers, the bit error rate may increases remarkably compared to linear power amplifiers \cite{Bj2013Impairment}. Although the deleterious impacts of hardware impairments on the security performance can be mitigated by calibration and compensation algorithms, residual distortions at the transceivers are inevitable \cite{Boshkovska2017Power}.

Many unknown challenges may be caused by hardware impairments in the fifth generation (5G) and beyond networks where novel physical layer technologies will be deployed, such as the technologies of massive MIMO, mm-Wave, and full duplex. In massive MIMO systems, additional challenges root in decreasing the hardware cost and increasing the power efficiency on antenna array which rise to hardware impairments. Moreover, due to the very large size of antenna array, standard algorithms for hardware impairment compensation, such as digital predistortion and phase-noise estimation and compensation may be too complex in a massive MIMO system \cite{Gust2014Onimpair}. The mm-Wave technologies utilize high frequency of mm-Wave band, ranging from $3\sim300$ GHz. Due to the very small wavelength, the mm-Wave networks are different from the conventional microwave networks in the following ways: large number of antennas, sensitivity to blockages, and variable propagation laws, which may deteriorate the harmful influence of hardware impairments to secure transmissions. In full duplex systems where the information is exchanged on the same frequency and time slot, the residual self-interference is still remained due to the impairments of hardware interference suppression methods, and signal processing technologies are needed to be addressed to suppress the residual self-interference thoroughly. In addition to those challenges mentioned above, in some infrastructureless networks and low-end networks (such as some specific scenarios in IoT) in which the communication equipments may be low-cost with small battery capacity, the hardware impairments may be more severe issues for implementing physical layer security.

\subsection{The Joint Designs of Physical Layer Security and Classic Cryptographic Security}
Some efforts may be needed for seeking deep insights into physical layer security and classic cryptographic security. In future, 5G network and beyond require ultra-strong security to support extremely secure service. Classic cryptographic security at the high cost of computational complexity, is usually deployed at the higher layers of protocol stack. As an alternative security technology, physical layer security has the advantages of lower complexity and resource savings. Any single security technology may not satisfy the demands of high security in future. Therefore, a natural question is how to jointly exploit the advantages of the two security technologies. Then, the cross-layer analysis and design combined with physical layer security and classic cryptographic security come naturally to mind to provide a comprehensive security solution from each layer of protocol stack.

To this end, there are many challenging problems needed to be solved in this direction, such as the secure network framework, secure coding scheme, secure network protocol, hybrid encryption algorithm, and so on. In future, the network architecture presents heterogeneous features, where the communication nodes are deployed with dissimilar characteristics such as computing capacity, energy supply capacity, radio access technologies, protocol stack architecture, etc. This requires that the joint security strategy designs can adapt to the heterogeneous architecture of networks, the variety of nodes, and the diversification of radio access technologies. This is significant but challenging work, since a joint security scheme for high level secrecy is usually followed with extremely high complexity which may limit its practical application. Moreover, the joint security scheme is expected to have a good scalability which allows the minimum amount of recomputation to update protocol parameters if some components of a network are changed. Therefore, in practice, how to design a simple but well-performing joint security scheme to tradeoff between the performance and the complexity is an urgent need to be addressed.

\subsection{The Global Optimization with Security, Reliability, and Throughput}

To achieve the optimal network performance and user experience in a wireless network, the security, reliability, and throughput should be considered jointly in system designs \cite{Zou2016security}. However, in many existing works, these performance metrics are taken into account individually and separately to reduce the difficulty in system designs. Consequently, the proposed security mechanisms are potentially suboptimal, since the three factors interact with each other. For instance \cite{Zou2016security}, the reliability and throughput of the legitimate channel can be improved by increasing the transmission power which however may improve the capacity of wiretap channel and increase the probability of successful eavesdropping. Likewise, although we can increase the coding rate at the transmitter for improving the security level while reducing the intercept probability, this leads to a decrease in transmission reliability, since higher coding rate may increase the outage probability of legitimate channel.

In order to achieve the near-perfect system performance, the global optimization with the joint considerations of security, reliability and throughput is needed to be carried out, which may be challenging and intractable. For formulating and solving such complicated multi-objective problems, some convex/nonconvex optimization techniques and game theory, as well as stochastic geometry, will be widely applied in this field \cite{Zhang2015Partner, Fakoorian2011MIMO}. Furthermore, the EE of a network attracts increasing concerns at present and in future. When the requirement of EE is imposed on the global optimization discussed above, the secure transmission designs will be extremely complicated work which calls for innovative efforts to develop novel optimization theories and technologies.

\subsection{The Commercial Application of Physical Layer Security}
It is largely unexplored to apply the technologies of physical layer security into commercial wireless networks. In fact, the most research work on physical layer security still stays at the theory stage. The opportunities of applying physical layer security into real commercial networks will be quite rich while following numerous difficulties and challenges that are from not only the technical flaws of the proposed secure strategies but also the limitations of existing network architecture and technologies, such as the hurdles from the applicability of existing network framework, the expansibility of underlying air interface, and the constraints of network resources \cite{Mukh2014Principles}.

Some new technical challenges will also be raised when physical layer security are applied into the burgeoning wireless networks, such as high-speed mobile networks, device to device communications, cognitive radio networks, and IoT. For example, in high-speed mobile networks as representative Internet of Vehicles and railway communication systems, the rapid changes of wireless channels and terminal positions require to propose fast CSI evaluation schemes and dynamic authentication frameworks. In device to device communications, due to direct communications between two mobile users without the supports of base stations or core networks, it is more difficult to establish a secure and reliable connection. Cognitive radio technique, as a promising technique to alleviate spectrum scarcity, has inherent vulnerabilities in physical layer spectrum sensing, such as the harmful interference from secondary users and the impersonation attack of disguised secondary users. To detect the disguised secondary users and to mitigate secondary interference, the terminals in cognitive radio networks should have the ability of autonomous learning. Machine learning is a powerful tool that can bring inspirations to cope with the potential challenges. IoT has a lot of particular characteristics, such as a massive number of devices, low-cost hardware, limited battery capacity, weak computation ability, and distinct service scenarios, all of which bring unprecedented challenges in implementing physical layer security.

\section{Conclusions} \label{section_conclusion}

It is believed that physical layer security is a promising technology to strengthen the secrecy of confidential information delivery in many emerging wireless networks in which the information security has not been well solved by the conventional cryptographic methods. To understand the advantages of physical layer security, a comparison is first made between this security technology and the conventional cryptographic encryption. Then, the survey mainly focuses on providing a comprehensive overview on the optimization and design of physical-layer security transmission. The typical wiretap channel models are introduced to cover common scenarios and systems in physical layer security. The research topics in this field are summarized from secure resource allocation, beamforming/precoding, and antenna/node selection and cooperation. Towards these research topics, we then discuss the performance metrics and fundamental optimization problems raised in the system optimization and design, which involve the secrecy rate/capacity, secrecy outage probability/capacity, power/energy consumption, and secure EE. The practical significance and applied scenarios of the metrics are also investigated in the survey. Each research topic of physical-layer security designs involves using these performance metrics to formulate optimization problems according to specific application conditions. Thereafter, the state of the art of optimization and design in physical layer security is reviewed from the perspectives of the aforementioned research topics. In each research topic, the great efforts are presented from four categories of fundamental optimization problems, such as maximization of achievable secrecy rate, minimization of secrecy outrage probability, minimization of power consumption, and maximization of secure EE. Numerous optimization approaches and solution schemes are investigated in the survey to tackle different problems in security designs.

One of the major issues in the physical-layer security designs is the imperfect CSI problem. To achieve the optimal performance of system designs, the transmitters need to know the CSI of both the legitimate users and the eavesdroppers. However, in practice, getting the perfect CSI of the eavesdroppers is very difficult or even impossible. This problem exists in all research topics of physical-layer security designs. In the survey, we review the existing assumptions of CSI which have been considered in physical layer security, while we discuss three ways to characterize the uncertainties of the imperfect eavesdropper's CSI. It is observed that, to cope with the problems of the imperfect or unknown CSI of eavesdroppers, the robust security designs, probabilistic view of security, or QoS-based optimization is usually considered in physical layer security to get a compromise solution. In addition, we discuss possible future trends and open challenges from the aspects involving the problems of imperfect CSI, eavesdropper models, and hardware impairments, as well as cross-layer security designs, global performance optimizations, and commercial application of physical layer security.

\bibliographystyle{IEEEtran}
\bibliography{Tutorials_reference}

\end{document}